\documentclass[12pt]{article}
\newcommand{\nc}{\newcommand}
\nc{\renc}{\renewcommand}

\usepackage{calc}
\usepackage{ifthen}
\usepackage{graphics}
\usepackage{epsfig}
\usepackage{miniplot}

\usepackage{wrapfig}
\usepackage{subfigure}
\usepackage{rotfloat}

\usepackage{flafter}

\nc{\half}{{\textstyle{1\over2}}}
\nc{\etal}{\mbox{\it et al. }}
\nc{\ie}{{\it i.e.}}
\nc{\eg}{{\it e.g.}}

\renc{\thefootnote}{\arabic{footnote}}
\nc{\capt}[1]{{\bf Figure.} {\small\sl #1}}

\nc{\eqs}[2]{\mbox{Eqs.~(\ref{#1},\,\ref{#2})}}
\nc{\eq}[1]{\mbox{Eq.~(\ref{#1})}}

\nc{\figs}[2]{\mbox{Figs.~(\ref{#1},\,\ref{#2})}}
\nc{\fig}[1]{\mbox{Fig~.(\ref{#1})}}

\nc{\tag}[1]{\label{#1} \marginpar{{\footnotesize #1}}}
\nc{\mtag}[1]{\label{#1} \mbox{\marginpar{{\footnotesize #1}}}}
\renc{\baselinestretch}{1.5}
\newlength{\overeqskip}
\newlength{\undereqskip}
\nc{\be}[1]{\begin{equation} \mbox{$\label{#1}$}}
\nc{\bea}[1]{\begin{eqnarray} \mbox{$\label{#1}$}}
\nc{\Section}[2]{\section{#2}\label{#1}}
\nc{\Bibitem}[1]{\bibitem{#1}}
\nc{\Label}[1]{\label{#1}}

\nc{\eea}{\vspace{\undereqskip}\end{eqnarray}}
\nc{\ee}{\vspace{\undereqskip}\end{equation}}
\nc{\bdm}{\begin{displaymath}}
\nc{\edm}{\end{displaymath}}
\nc{\dpsty}{\displaystyle}
\nc{\bc}{\begin{center}}
\nc{\ec}{\end{center}}
\nc{\ba}{\begin{array}}
\nc{\ea}{\end{array}}
\nc{\bab}{\begin{abstract}}
\nc{\eab}{\end{abstract}}
\nc{\btab}{\begin{tabular}}
\nc{\etab}{\end{tabular}}
\nc{\bit}{\begin{itemize}}
\nc{\eit}{\end{itemize}}
\nc{\ben}{\begin{enumerate}}
\nc{\een}{\end{enumerate}}
\nc{\bfig}{\begin{figure}}
\nc{\efig}{\end{figure}}
\nc{\arreq}{&\!=\!&}
\nc{\arrmi}{&\!-\!&}
\nc{\arrpl}{&\!+\!&}
\nc{\arrap}{&\!\!\!\approx\!\!\!&}
\nc{\non}{\nonumber\\*}
\nc{\align}{\!\!\!\!\!\!\!\!&&}

\def\lsim{\; \raise0.3ex\hbox{$<$\kern-0.75em
      \raise-1.1ex\hbox{$\sim$}}\; }
\def\gsim{\; \raise0.3ex\hbox{$>$\kern-0.75em
      \raise-1.1ex\hbox{$\sim$}}\; }
\nc{\DOT}{\hspace{-0.08in}{\bf .}\hspace{0.1in}}
\nc{\Laada}{\hbox {$\sqcap$ \kern -1em $\sqcup$}}
\nc\loota{{\scriptstyle\sqcap\kern-0.55em\hbox{$\scriptstyle\sqcup$}}}
\nc\Loota{{\sqcap\kern-0.65em\hbox{$\sqcup$}}}
\nc\laada{\Loota}
\nc{\qed}{\hskip 3em \hbox{\BOX} \vskip 2ex}

\nc{\real}{{\rm I \! R}}
\nc{\Z}{{\sf Z \!\!\! Z}}
\nc{\complex}{{\rm C\!\!\! {\sf I}\,\,}}
\def\bigid{\leavevmode\hbox{\small1\kern-3.8pt\normalsize1}}
\def\id{\leavevmode\hbox{\small1\kern-3.3pt\normalsize1}}
\nc{\slask}{\!\!\!/}
\nc{\bis}{{\prime\prime}}
\nc{\pa}{\partial}
\nc{\na}{\nabla}
\nc{\ra}{\rangle}
\nc{\la}{\langle}
\nc{\goto}{\rightarrow}
\nc{\swap}{\leftrightarrow}

\nc{\EE}[1]{ \mbox{$\cdot10^{#1}$} }
\nc{\abs}[1]{\left|#1\right|}
\nc{\at}[2]{\left.#1\right|_{#2}}
\nc{\norm}[1]{\|#1\|}
\nc{\abscut}[2]{\Abs{#1}_{\scriptscriptstyle#2}}
\nc{\vek}[1]{{\rm\bf #1}}
\nc{\integral}[2]{\int\limits_{#1}^{#2}}
\nc{\inv}[1]{\frac{1}{#1}}
\nc{\dd}[2]{{{\partial #1}\over{\partial #2}}}
\nc{\ddd}[2]{{{{\partial}^2 #1}\over{\partial {#2}^2}}}
\nc{\dddd}[3]{{{{\partial}^2 #1}\over
        {\partial #2 \partial #3}}}
\nc{\dder}[2]{{{d #1}\over{d #2}}}
\nc{\ddder}[2]{{{d^2 #1}\over{d {#2}^2}}}
\nc{\dddder}[3]{{d^2 #1}\over
        {d #2 d #3}}
\nc{\dx}[1]{d\,^{#1}x}
\nc{\dy}[1]{d\,^{#1}y}
\nc{\dz}[1]{d\,^{#1}z}
\nc{\dl}[1]{\frac{d\,^{#1}l}{(2\pi)^{#1}}}
\nc{\dk}[1]{\frac{d\,^{#1}k}{(2\pi)^{#1}}}
\nc{\dq}[1]{\frac{d\,^{#1}q}{(2\pi)^{#1}}}

\nc{\cc}{\mbox{$c.c.$ }}
\nc{\hc}{\mbox{$h.c.$ }}
\nc{\cf}{cf.\ }
\nc{\erfc}{{\rm erfc}}
\nc{\Tr}{{\rm Tr\,}}
\nc{\tr}{{\rm tr\,}}
\nc{\pol}{{\rm pol}}
\nc{\sign}{{\rm sign}}
\nc{\bfT}{{\bf T }}

\def\GeV{{\rm\ GeV}}

\nc{\cA}{{\cal A}}
\nc{\cB}{{\cal B}}
\nc{\cD}{{\cal D}}
\nc{\cE}{{\cal E}}
\nc{\cG}{{\cal G}}
\nc{\cH}{{\cal H}}
\nc{\cL}{{\cal L}}
\nc{\cO}{{\cal O}}
\nc{\cT}{{\cal T}}
\nc{\cN}{{\cal N}}
\nc{\rvac}[1]{|{\cal O}#1\rangle}
\nc{\lvac}[1]{\langle{\cal O}#1|}
\nc{\rvacb}[1]{|{\cal O}_\beta #1\rangle}
\nc{\lvacb}[1]{\langle{\cal O}_\beta #1 |}
\nc{\bb}{\bar{\beta}}
\nc{\bt}{\tilde{\beta}}
\nc{\ctH}{\tilde{\cal H}}
\nc{\chH}{\hat{\cal H}}
\nc{\1}{\aa}
\nc{\2}{\"{a}}
\nc{\3}{\"{o}}
\nc{\4}{\AA}
\nc{\5}{\"{A}}
\nc{\6}{\"{O}}
\nc{\al}{\alpha}
\nc{\g}{\gamma}
\nc{\Del}{\Delta}
\nc{\e}{\epsilon}
\nc{\eps}{\epsilon}
\nc{\lam}{\lambda}
\nc{\om}{\omega}
\nc{\Om}{\Omega}
\nc{\ve}{\varepsilon}
\nc{\mn}{{\mu\nu}}
\nc{\vp}{\varphi}

\nc{\advp}[3]{{\it  Adv.\ in\ Phys.\ }{{\bf #1} {(#2)} {#3}}}
\nc{\annp}[3]{{\it  Ann.\ Phys.\ (N.Y.)\ }{{\bf #1} {(#2)} {#3}}}
\nc{\apl}[3]{{\it  Appl. Phys. Lett. }{{\bf #1} {(#2)} {#3}}}
\nc{\apj}[3]{{\it  Ap.\ J.\ }{{\bf #1} {(#2)} {#3}}}
\nc{\apjl}[3]{{\it  Ap.\ J.\ Lett.\ }{{\bf #1} {(#2)} {#3}}}
\nc{\app}[3]{{\it Astropart.\ Phys.\ }{{\bf #1} {(#2)} {#3}}}
\nc{\cmp}[3]{{\it  Comm.\ Math.\ Phys.\ }{{ \bf #1} {(#2)} {#3}}}
\nc{\cqg}[3]{{\it  Class.\ Quant.\ Grav.\ }{{\bf #1} {(#2)} {#3}}}
\nc{\epl}[3]{{\it  Europhys.\ Lett.\ }{{\bf #1} {(#2)} {#3}}}
\nc{\ijmp}[3]{{\it Int.\ J.\ Mod.\ Phys.\ }{{\bf #1} {(#2)} {#3}}}
\nc{\ijtp}[3]{{\it Int.\ J.\ Theor.\ Phys.\ }{{\bf #1} {(#2)} {#3}}}
\nc{\jmp}[3]{{\it  J.\ Math.\ Phys.\ }{{ \bf #1} {(#2)} {#3}}}
\nc{\jpa}[3]{{\it  J.\ Phys.\ A\ }{{\bf #1} {(#2)} {#3}}}
\nc{\jpc}[3]{{\it  J.\ Phys.\ C\ }{{\bf #1} {(#2)} {#3}}}
\nc{\jap}[3]{{\it J.\ Appl.\ Phys.\ }{{\bf #1} {(#2)} {#3}}}
\nc{\jpsj}[3]{{\it J.\ Phys.\ Soc.\ Japan\ }{{\bf #1} {(#2)} {#3}}}
\nc{\lmp}[3]{{\it Lett.\ Math.\ Phys.\ }{{\bf #1} {(#2)} {#3}}}
\nc{\mpl}[3]{{\it  Mod.\ Phys.\ Lett.\ }{{\bf #1} {(#2)} {#3}}}
\nc{\ncim}[3]{{\it  Nuov.\ Cim.\ }{{\bf #1} {(#2)} {#3}}}
\nc{\np}[3]{{\it  Nucl.\ Phys.\ }{{\bf #1} {(#2)} {#3}}}
\nc{\npps}[3]{{\it  Nucl.\ Phys.\ Proc.\ Suppl.\ }{{\bf #1} {(#2)} {#3}}}
\nc{\pr}[3]{{\it Phys.\ Rev.\ }{{\bf #1} {(#2)} {#3}}}
\nc{\pra}[3]{{\it  Phys.\ Rev.\ A\ }{{\bf #1} {(#2)} {#3}}}
\nc{\prb}[3]{{\it  Phys.\ Rev.\ B\ }{{{\bf #1} {(#2)} {#3}}}}
\nc{\prc}[3]{{\it  Phys.\ Rev.\ C\ }{{\bf #1} {(#2)} {#3}}}
\nc{\prd}[3]{{\it  Phys.\ Rev.\ D\ }{{\bf #1} {(#2)} {#3}}}
\nc{\prl}[3]{{\it Phys.\ Rev.\ Lett.\ }{{\bf #1} {(#2)} {#3}}}
\nc{\pl}[3]{{\it  Phys.\ Lett.\ }{{\bf #1} {(#2)} {#3}}}
\nc{\prep}[3]{{\it Phys.\ Rep.\ }{{\bf #1} {(#2)} {#3}}}
\nc{\prsl}[3]{{\it Proc.\ R.\ Soc.\ London\ }{{\bf #1} {(#2)} {#3}}}
\nc{\ptp}[3]{{\it  Prog.\ Theor.\ Phys.\ }{{\bf #1} {(#2)} {#3}}}
\nc{\ptps}[3]{{\it  Prog\ Theor.\ Phys.\ suppl.\ }{{\bf #1} {(#2)} {#3}}}
\nc{\physa}[3]{{\it  Physica\ A\ }{{\bf #1} {(#2)} {#3}}}
\nc{\physb}[3]{{\it  Physica\ B\ }{{\bf #1} {(#2)} {#3}}}
\nc{\phys}[3]{{\it Physica\ }{{\bf #1} {(#2)} {#3}}}
\nc{\rmp}[3]{{\it  Rev.\ Mod.\ Phys.\ }{{\bf #1} {(#2)} {#3}}}
\nc{\rpp}[3]{{\it Rep.\ Prog.\ Phys.\ }{{\bf #1} {(#2)} {#3}}}
\nc{\sjnp}[3]{{\it Sov.\ J.\ Nucl.\ Phys.\ }{{\bf #1} {(#2)} {#3}}}
\nc{\spjetp}[3]{{\it Sov.\ Phys.\ JETP\ }{{\bf #1} {(#2)} {#3}}}
\nc{\yf}[3]{{\it Yad.\ Fiz.\ }{{\bf #1} {(#2)} {#3}}}
\nc{\zetp}[3]{{\it Zh.\ Eksp.\ Teor.\ Fiz.\  }{{\bf #1}  {(#2)} {#3}}}
\nc{\zp}[3]{{\it Z.\ Phys.\ }{{\bf #1} {(#2)} {#3}}}
\nc{\ibid}[3]{{\sl ibid.\ }{{\bf #1} {#2} {#3}}}
\nc{\rf}[1]{(\ref{#1})}
\nc{\nn}{\nonumber \\*}
\nc{\bfB}{\bf{B}}
\nc{\bfv}{\bf{v}}
\nc{\bfx}{\bf{x}}
\nc{\bfy}{\bf{y}}
\nc{\vx}{\vec{x}}
\nc{\vy}{\vec{y}}
\nc{\oB}{\overline{B}}
\nc{\oI}{\overline{I}}
\nc{\oR}{\overline{R}}
\nc{\rar}{\rightarrow}
\nc{\ti}{\times}
\nc{\slsh}{\hskip-5pt/}
\nc{\sm}{Standard~Model~}
\nc{\MP}{M_{\rm Pl}}
\nc{\tp}{t_{\rm Pl}}
\nc{\ave}{\bar{E}}

\nc{\eff}{{\rm eff}}
\nc{\kk}{\vek{k}}
\nc{\pp}{{\rm p}}
\nc{\ga}{g_{a\gamma}}
\nc{\vv}{\\}
\nc{\eee}{{\bf E}}
\nc{\bbb}{{\bf B}}
\nc{\qcd}{T_{\rm QCD}}
\nc{\G}{\rm \ G}
\def\vec#1{{\bf #1}}

\def\lae{\;^{<}_{\sim} \;} \def\gae{\; ^{>}_{\sim} \;} 

\def\ell{e^{c}LL}

\begin{document}
{\title{\vskip-2truecm{\hfill {{\small \\
	\hfill \\
	}}\vskip 1truecm}
{\bf Growth of Inflaton Perturbations 
and the Post-Inflation Era  
in Supersymmetric Hybrid Inflation Models}}
{\author{
{\sc  Matt Broadhead$^{1}$ and John McDonald$^{2}$}\\
{\sl\small Dept. of Mathematical Sciences, University of Liverpool,
Liverpool L69 3BX, England}
}
\maketitle
\begin{abstract}
\noindent

    It has been shown
that hybrid inflation may end with the formation of non-topological 
solitons of inflaton field. 
As a first step towards a fully realistic picture of the post-inflation 
era and reheating in supersymmetric hybrid inflation models, 
we study the classical scalar field equations 
of a supersymmetric hybrid inflation model using
a semi-analytical ansatz for the spatial dependence of the fields. 
Using the minimal D-term inflation model as an example,
 the inflaton field is evolved using the 
full 1-loop effective potential from the
 slow-rolling era to the $U(1)_{FI}$ symmetry-breaking phase transition.
 Spatial perturbations of the  
inflaton corresponding to quantum fluctuations are introduced
for the case where there is spatially coherent $U(1)_{FI}$ symmetry breaking. 
The maximal growth of the dominant perturbation is found to depend only
 on the ratio of superpotential coupling 
$\lambda$ to the gauge coupling $g$.  The inflaton condensate 
fragments to non-topological solitons for $\lambda/g \gae 0.09$.  
Possible consequences of non-topological soliton formation 
in fully realistic SUSY hybrid inflation models are discussed.

\end{abstract}
\vfil
\footnoterule
{\small $^1$mattb@amtp.liv.ac.uk}
{\small $^2$mcdonald@amtp.liv.ac.uk}

\thispagestyle{empty}
\newpage
\setcounter{page}{1}

\section{Introduction}

         An important event in the history of the Universe is the process by
 which it became hot after inflation, known as reheating. A full understanding of this
 epoch is essential for a full understanding of cosmology. It is also important to have 
a clear picture of the post-inflation cosmological 
enviroment against which ideas in particle physics and cosmology can be
judged. The conventional view of reheating following inflation has been based on the
perturbative decay of a spatially homogeneous Bose condensate of inflaton scalars
 \cite{eu}.
This picture, however, has been challanged in recent years. It was firstly observed that 
the inflaton condensate might decay non-perturbatively via parametric resonance
 \cite{param0,param1,param2}. 
More recently, for some inflation models, in particular hybrid inflation \cite{hi}, it was shown 
that the inflaton sector Bose condensate can be unstable with respect to spatial
perturbations of the scalar fields \cite{icf,tp}. The growth of the spatial perturbations 
could either
be slow, in which case a coherently oscillating condensate would form prior to spatial
perturbations going non-linear ('inflaton condensate fragmentation') \cite{icf}, or
 rapid, in which
case the spatial perturbations dominate the total energy density before the inflaton can
enter into coherent oscillations ('tachyonic preheating') \cite{tp}. In the case of inflaton
condensate fragmentation it was observed that the final state of the process is likely
 to be, at least for some range of parameters of the model, 
the formation of non-topological soliton-like objects 
('inflaton condensate lumps') \cite{icf,iballs}.
 The formation of such objects in hybrid inflation models
was subsequently confirmed in lattice simulations \cite{cpr}. Recently it has been 
shown that inflaton condensate fragmentation can also occur in single field 
chaotic inflation models \cite{keq}.

         However, there has been to date no detailed 
study of condensate fragmentation and non-topological soliton formation in a 
fully realistic hybrid inflation model. Of particular interest is the case of 
supersymmetric (SUSY) hybrid inflation models, which are prime candidates for 
natural inflation models in the context of SUSY \cite{fti,dti}. (The 
alternative, SUSY chaotic 
inflation, is disfavoured by the initial inflaton expectation value 
required, larger than 
the natural scale of supergravity (SUGRA) 
corrections, $M = M_{Pl}/\sqrt{8 \pi}$ 
\cite{chaotic}.)  
In the following we will focus on the case of the minimal D-term hybrid inflation
 model 
as a specific example. In D-term inflation 
models the energy density is provided by a Fayet-Illiopoulos D-term associated with a 
$U(1)_{FI}$ gauge symmetry \cite{dti}. D-term inflation 
is particularly favoured as it provides a natural solution to the $\eta$-problem 
of SUGRA inflation models \cite{eta}. 

           Hybrid inflation models in general are based on two
 scalar fields, an inflaton, $S$, and a second field, $\Phi$, which undergoes a 
symmetry-breaking phase transition which terminates inflation. (We will refer to
 these 
as the 'inflaton sector' fields.) As well as the homogeneous scalar fields considered in 
conventional D-term inflation models, there will also be quantum 
fluctuations leading to spatial perturbations of the 
scalar fields \cite{eu}. 
In order to establish the initial conditions for the growth of spatial 
perturbations following inflation, it
is essential to follow the evolution of the inflaton 
from the slow-rolling epoch during inflation until the phase
transition which terminates hybrid inflation. 
The evolution of the inflaton prior to the phase transition in SUSY hybrid inflation
 models 
is dominated by the 1-loop effective potential. 
Therefore in the following we will consider the evolution of the inflaton sector fields 
in D-term inflation including the full 1-loop radiative correction.

          Spatial perturbations of the inflaton sector fields arise initially as a result of
 quantum fluctuations. A method to study the evolution of these fluctuations into the
 classical regime has been developed in \cite{ps}. 
So long as the final state of the evolution of the scalar fields is
 semi-classical, the evolution may be studied by solving the classical equations of 
motion 
but with a classical probability distribution for the initial conditions, determined by
 the 
initial quantum state of the field. (Following \cite{ps}, we will 
refer to this as the 'equivalent classical
 stochastic field' (ECSF) method in the following.) 

  In this paper we study the evolution of spatial perturbations in D-term inflation 
using a semi-analytical model with simplified initial conditions for the spatial
 perturbations. This will allow us to gain a physical understanding of the 
process of spatial perturbation growth whilst providing an example against which 
future numerical simulations may be tested. 
We will 
argue that the growth of perturbations of the inflaton in the 
semi-analytical model represents 
the maximum possible growth in realistic models. In addition, we will 
show that that the formation of cosmic strings is likely to be 
rare on scales smaller than 
the horizon and so is unlikely to play a significant role in the subsequent evolution of 
the energy density, in which case the model we consider should give a good estimate 
of the growth of perturbations in the realistic case. 

           As in all previous studies, we will consider 
the evolution of a real inflaton field. 
However, in fully realistic models of SUSY inflation, 
the inflaton is a complex field. As such, none of 
the existing analyses of spatial 
perturbation growth in hybrid inflation models 
fully describes the SUSY case. One result of 
the complex nature of the inflaton field is that the non-topological solitons 
which form at the end of hybrid inflation are likely to be Q-balls \cite{keq,mattq}. 
We will discuss how the analysis presented here relates to Q-ball formation and 
discuss some possible cosmological consequences of non-topological soliton 
formation in SUSY hybrid inflation models. 

       The paper is organised as follows. In Section 2 we discuss the D-term 
inflation model and the full 1-loop effective potential. In Section 3 we discuss the 
initial conditions used for the spatial perturbations. 
In Section 4 we discuss the semi-analytical ansatz used to model the growth of 
spatial perturbations. In Section 5 we discuss the evolution of perturbations of the 
inflaton sector fields in the semi-analytical model.
In Section 6 we discuss some possible consequences of 
non-topological soliton formation in realistic SUSY hybrid inflation models. 
In Section 7 we present our conclusions. 
\newpage

\section{Equations of Motion and 1-loop Effective Potential}

     In the following we will consider the minimal SUSY 
D-term hybrid inflation model\footnote{The tree-level scalar potential 
for the D-term inflation model becomes equivalent to the minimal 
F-term inflation model \cite{fti} in the limit where 
$\lambda = \sqrt{2}g$. Therefore study of the D-term model should  
also give an insight into the growth of spatial perturbations in the 
F-term inflation model.} \cite{dti}. 
The superpotential of the minimal D-term inflation model is  
\be{e1} W = \lambda S \Phi_{+} \Phi_{-}    ~.\ee
    The tree-level scalar potential is then 
\be{e2} V = \lambda^{2}|S|^{2}\left(|\Phi_{+}|^{2} + |\Phi_{-}|^{2}\right) 
+ \lambda^{2} |\Phi_{+}|^{2} |\Phi_{-}|^{2} + \frac{g^{2}}{2} \left(|\Phi_{+}|^{2}
- |\Phi_{-}|^{2} + \xi\right)^{2}    ~,\ee
where $S$ is the inflaton, $\Phi_{\pm}$ are fields with charges $\pm$1 with respect
 to a 
Fayet-Illiopoulos $U(1)_{FI}$ gauge symmetry, $\xi > 0$ is the Fayet-Illiopoulos
 gauge term ($\xi^{1/2} \approx 8.5 \times 10^{15}\GeV$ \cite{lr}) and 
$g$ is the $U(1)_{FI}$ gauge coupling. Since the minimum of the potential as a function of $S$ 
is generally at  
$|\Phi_{+}| = 0$ and since $|\Phi_{+}| = 0$ at the end of inflation, we 
will assume 
this value throughout. (We will discuss this assumption later.) We will 
also consider the
 scalar potential along 
the real 
$S$ and $\Phi_{-}$ directions. Thus with $s= \sqrt{2} Re(S)$ and $\phi_{-} =
 \sqrt{2}Re(\Phi_{-})$, the 
tree-level scalar potential becomes
\be{e2} V(s,\phi_{-}) = \frac{\lambda^{2}}{4} s^{2} \phi_{-}^{2}  + \frac{g^{2}}{2} \left(\xi 
- \frac{\phi_{-}^{2}}{2}\right)^{2}   ~.\ee
The $U(1)_{FI}$ symmetry breaking transition occurs once $s < s_{c} \equiv
 \sqrt{2} g \xi^{1/2}/\lambda$.

         In addition, we require the 1-loop
 effective potential as a function of $s$ and $\phi_{-}$, 
$\Delta V(s,\phi_{-})$. This has been done previously for the case 
where $\phi_{-}$ is fixed at the minimum of its potential as a 
function of $s$ \cite{jeannerot}. 
However, as we wish to discuss the equations of motion 
of the $s$ and $\phi_{-}$ field seperately here, 
we require the mass eigenvalues and 
effective potential as a function of both fields. For $s < s_{c}$ the 
scalar and gauge boson mass terms and degrees of freedom 
as a function of $s$ and $\phi_{-}$ are given in Table 1.
\newline

\begin{table}[htbp]
\begin{center}
\begin{tabular}{|c|c|c|} \hline Field & d.o.f & Mass Squared \\
\hline $\eta_{+}$ & 1 & $\frac{1}{2} \left[ B + \sqrt{B^{2} - 4C}\right]$ \\
$\eta_{-}$ & 1 & $\frac{1}{2} \left[ B - \sqrt{B^{2} - 4C}\right]$ \\
$s_{2}$ & 1 & $\frac{\lambda^{2}}{2} \phi_{-}^{2}$    \\ 
$\Phi_{+}$ & 2 & $\frac{\lambda^{2}}{2} s^{2} 
+ \left(\frac{\lambda^{2} -g^{2}}{2}\right)\phi_{-}^{2} + g^{2} \xi$
\\ 
$A$ & 3 & $g^{2}\phi_{-}^{2}$ \\
\hline     

\end{tabular}
\caption{\footnotesize{Mass terms and physical degrees of freedom for $s<s_c$.}}  

\end{center}
\end{table}

       Here $B = \frac{1}{2}\lambda^{2} \left(s^{2} + \phi_{-}^{2}\right) 
+ \frac{3 }{2}g^{2}\phi_{-}^{2} - g^{2} \xi$ and 
$C = \frac{1 }{2}\lambda^{2} \phi_{-}^{2} \left(\frac{1 }{2}\lambda^{2}s^{2}
+ \frac{3 }{2}g^{2} \phi_{-}^{2} - g^{2} \xi \right) 
- \frac{1}{4}\lambda^{4} s^{2} \phi_{-}^{2}$. $\eta_{\pm}$ are the eigenstates of
 the 
$s_{1}$, $\phi_{1}$ mass matrix. (We define $S = (s_{1} + i s_{2})/\sqrt{2}$ and 
$\Phi_{-} = (\phi_{1} + i \phi_{2})/\sqrt{2}$.) 
$A$ is the $U(1)_{FI}$ 
gauge boson, with $\phi_{2}$ the corresponding Goldstone boson\footnote{More
 precisely, 
$\Phi_{-} = \frac{\phi}{\sqrt{2}}e^{i \theta}$, 
with $\theta$ the Goldstone boson degree of freedom. 
The mass terms of the physical scalar field $\phi$ with $\theta = 0$ are the same as those of 
 $\phi_{1}$ with $\phi_{2} = 0$.}. 

   The fermion mass eigenstates come from the mass matrix $M_{\Psi}$ for 
$\Psi \equiv (\chi, \lambda_{+}, \lambda_{-}, \lambda_{S})$, where $\chi$ is the 
$U(1)_{FI}$ gaugino, $\lambda_{+}$($\lambda_{-}$) 
is the fermionic component of 
$\Phi_{+}$($\Phi_{-}$) and $\lambda_{S}$ is the 
fermionic component of the $S$ chiral superfield, 
\be{e3} \Psi^{T}M_{\Psi}\Psi \equiv \left( \begin{array}{cccc} \chi & \lambda_{-} 
& \lambda_{+} & \lambda_{S} \\ 
\end{array} \right) 
\left[ \begin{array}{cccc} 
0 & g \phi_{-} & 0 & 0 \\
g \phi_{-} & 0 & \frac{\lambda}{\sqrt{2}}s& 0 \\ 
0 & \frac{\lambda}{\sqrt{2}}s & 0 & \frac{\lambda}{\sqrt{2}}\phi_{-} \\
0 & 0 & \frac{\lambda}{\sqrt{2}}\phi_{-} & 0 \\ 
\end{array}\right] \left( 
\begin{array}{c} \chi \\ \lambda_{-} \\ \lambda_{+} 
\\ \lambda_{S} \\  \end{array} \right)    ~.\ee
The resulting eigenstates correspond to two Dirac fermons, of mass 
 \be{e4}  m_{1,2} = \frac{1}{2}\left[ 
\frac{\lambda^{2}}{2} s^{2} + \left(g^{2} + 
\frac{\lambda^{2}}{2}\right) \phi_{-}^{2}\right] 
\pm \sqrt{\Delta}  ~,\ee
where 
\be{e5} \Delta = \frac{1}{4} \left(
\left[ \left( g^{2} + \frac{\lambda^{2}}{2} \right) \phi_{-}^{2} + 
\frac{\lambda^{2}}{2}s^{2} \right]^{2} - 2 \lambda^{2} g^{2} \phi^{4}\right)  ~.\ee

      For $s > s_{c}$ there are two massive complex scalars $\Phi_{\pm}$ and a single 
Dirac fermion $\Psi$, with squared masses given in Table 2.
\newline 

\begin{table}[htbp]
\begin{center}
\begin{tabular}{|c|c|c|} \hline Field & d.o.f & Mass Squared \\
\hline $\Phi_{+}$ & 2 & $ \frac{\lambda^{2}}{2}s^{2} + g^{2} \xi$ \\
$\Phi_{-}$ & 2 & $ \frac{\lambda^{2}}{2}s^{2} - g^{2} \xi $ \\
$\Psi$ & 4 & $ \frac{\lambda^{2}}{2}s^{2}$ \\
\hline     
\end{tabular}
\caption{\footnotesize{Mass terms and physical degrees of freedom for $s>s_c$.}}
\end{center}
\end{table}

     The 1-loop effective potential as a function of $s$ and $\phi$ is given by the 
Coleman-Weinberg formula 
\be{e6}   \Delta V(s, \phi) = \frac{1}{64 \pi^{2}} \sum_{i} \left(-1\right)^{F}m_{i}^{4} ln\left(
\frac{m_{i}^{2}}{\Lambda^{2}} \right)      ~,\ee
where $\Lambda$ is a renormalisation scale and 
the contribution is negative for fermions. An important 
question is how this should be interpreted 
when a scalar mass squared becomes negative for small field expectation values. This
 problem also arises 
in the case of the Standard Model Higgs effective potential. Weinberg and Wu
 \cite{weinberg} argue that 
only the real part of the effective potential corresponds to a correction to the energy
 density and so 
enters the equations of motion. Thus if $m_{i}^{2} < 0$, we can simply substitute
 $m_{i}^{2} = 
|m_{i}^{2}|e^{i \pi}$ in \eq{e6}, in which case the real part of $\Delta V$ corresponds to 
replacing $m_{i}^{2}$ by $|m_{i}^{2}|$ throughout. We will use this proceedure in the following.   

           The full equations of motion to 1-loop are then 
\be{e7} \ddot{s} + 3 H \dot{s} - \frac{\vec{\nabla}^{2}}{a^{2}}s    
= -\frac{\lambda^{2} \phi_{-}^{2}s}{2} - \frac{\partial \Delta V}{\partial s}      ~,\ee
and
\be{e8} \ddot{\phi}_{-} + 3 H \dot{\phi}_{-} - \frac{\vec{\nabla}^{2}}{a^{2}}\phi_{-}    
= -\frac{\lambda^{2} s^{2}\phi_{-}}{2} + g^{2}\left(\xi - \frac{\phi_{-}^{2}}{2} \right)
\phi_{-} - \frac{\partial \Delta V}{\partial \phi_{-}}      ~.\ee

\section{Spatial Perturbations and Initial Conditions} 

\subsection{Equivalent Classical Stochastic Field}

         We are interested in the growth of spatial perturbations of the scalar fields. 
These have their origin in quantum fluctuations. In studying the growth of quantum 
fluctuations of a scalar field, the method 
used is that of the equivalent classical 
stochastic field (ECSF) \cite{ps}. 
This is based on the observation that for quantum states which
 evolve to semi-classical final states, 
expectation values in the full quantum field theory 
are the same as those obtained by solving the classical equations
of motion with initial conditions which have a classical stochastic 
distribution determined by the 
initial quantum state\footnote{A simple and useful illustration of 
this concept in the context of quantum mechanics is given in \cite{guthpi}.}. 
 For the case we are considering here, the 
initial state is taken to correspond to the conformal vacuum state of a 
massless scalar field $\phi$ in (quasi) de Sitter space ($H \approx$ constant).  
The corresponding initial conditions for the field modes are \cite{ps}
\be{ic1} |y(\vec{k})|^{2} = \frac{1}{2k} \left(1 + \frac{1}{k^{2}\eta^{2}}\right) 
~,\ee
\be{ic2} |p(\vec{k})|^{2} = 
\left(\frac{1}{4 k^{2} \eta^{2}}\right) \frac{1}{|y(\vec{k})|^{2}}
= \frac{1}{2 k \eta^{2}} 
\left(1 + \frac{1}{k^{2} \eta^{2}}\right)^{-1}     ~.\ee
Here we have used the notation of \cite{ps}, where $y(\vec{k})$ is the 
Fourier transform of the perturbation of the scalar field $\phi(\vec{x})$, 
$\delta \phi(\vec{x}) \equiv y(\vec{x})/a $
 (time dependence not explicitly 
stated), and we are considering initial 
conditions taken at scale factor $a = 1$. $p(\vec{k})$ is the 
Fourier transform of $p(\vec{x})$, the conjugate variable of $y(\vec{x})$
\be{ic3}  p(\vec{x}) = y^{'} - \left(\frac{a^{'}}{a}\right)y     ~,\ee
where a prime denotes differentiation with respect to conformal time $\eta$ ($\eta =
 \int dt/a = -(aH)^{-1}$).  
These modes refer to Fourier transforms defined by 
\be{ic4}   y(\vec{x}) = \frac{V^{1/2}}{\left(2 \pi \right)^{3/2}}
 \int y(\vec{k}) e^{-i\vec{k}.\vec{x}} d^{3}k     ~,\ee
\be{ic5}   y(\vec{k}) = 
\frac{V^{-1/2}}{\left(2 \pi \right)^{3/2}} 
\int_{V} y(\vec{x}) e^{i\vec{k}.\vec{x}} d^{3}x     ~,\ee
where we retain an explicit spatial volume $V$ in order to 
account for the mass dimensions. For a real $y(\vec{x})$ we have 
$y(\vec{k}) = y^{*}(-\vec{k})$. 

          Given the initial conditions, the subsequent 
evolution of the system is obtained by evolving the scalar field classically. 
In practice the important quantities are the 
root mean square (r.m.s.) spatial averages of
the field fluctuations. For example, once the 
r.m.s. fluctuation of the field amplitude is of the order
of the homogeneous field amplitude, the field will almost entirely exist in the
overdense regions and therefore we expect the spatial perturbations to go non-linear and 
to fragment to non-topological solitons. The mean squared fluctuation is 
given by 
\be{ic6} \left<y\left(\vec{x}\right)^{2}\right> = 4 \pi \int k^{3} |y(\vec{k})|^{2} 
 \frac{dk}{k}     ~,\ee
where $k = |\vec{k}|$. 
Thus modes with $|\vec{k}|$ in a range $\Delta k \sim k$ around $k$ will contribute 
to the r.m.s. fluctuation squared an amount \cite{eu} 
\be{ic7}  \Delta_{k}^{2} \equiv \left<y\left(\vec{x}\right)^{2}\right>_{k} 
\approx 4 \pi k^{3} |y(\vec{k})|^{2}   ~.\ee
We will model the classical evolution of the r.m.s. 
fluctuation due to modes with $|\vec{k}| \sim k$ 
by considering a single mode with $|\vec{k}| = k$ and initial amplitude 
given by $\Delta_{k}$.
We will be particularly interested in the dominant 
mode (the largest at late times), since its 
wavelength is likely to determine the size of 
the non-topological solitons which form
 once the inflaton sector energy density perturbations 
become non-linear \cite{icf,iballs}. (This is similar to 
what is observed numerically in 
the case of a complex scalar condensate 
and Q-ball formation \cite{bbb,kk}.) 
The modes of interest in the following will have  
wavelength small compared with the horizon at the end of inflation. 
In this case $|k\eta|
 = k/H \gg 1$ and 
$|y(\vec{k})| \approx 1/\sqrt{2k}$. Then $\Delta_{k} \approx  \sqrt{2 \pi} k$.

   The momentum conjugate to $y(\vec{x})$ in the limit $k \gg H$ is given by 
\be{ic9}  p(\vec{x}) = a \left(\dot{y}(\vec{x}) -  H y(\vec{x})\right)  ~,\ee
where 'dot' denotes differentiation with respect to time. 
Thus if $\dot{y}(\vec{x})/y(\vec{x}) 
\equiv \delta\dot{\phi}(\vec{x})/\delta \phi(\vec{x}) \gae 
H$ (as expected if perturbation growth is significant) we have 
initially (at $a = 1$) $p(\vec{x}) 
\approx \dot{y}(\vec{x}) \equiv \delta \dot{\phi}(\vec{x})$.  
Therefore the initial mean squared rate of change of $\delta \phi(\vec{x})$ 
corresponding to the range
$\Delta k \sim k$ around $k$ is related to $p(\vec{k})$ by (cf. \eq{ic7}), 
\be{ic10}
 \left<\delta \dot{\phi}^{2}(\vec{x})\right>_{k} \approx 4 \pi k^{3} |p(\vec{k})|^{2}
=  2 \pi k^{4} \left(\frac{H}{k}\right)^{2}  ~.\ee
 
   Thus to model the ECSF initial conditions we will consider the evolution of a 
scalar field mode $\delta \phi_{k}(\vec{x})$ with initially (at $t = t_{o}$)  
\be{ic8} \delta \phi_{k}(\vec{x},t_{o})  =  A(t_{o}) \sin(\vec{k}.\vec{x}) ~,\ee
where 
\be{ic11} A(t_{o}) = \sqrt{2 \pi} k   ~\ee
and
\be{ic12} |\dot{A}(t_{o})| = \sqrt{2 \pi} k^{2}\left(\frac{H}{k}\right)    ~.\ee
(In practice we find that $|\dot{A}(t_{o})|$ has a negligible effect.) 

       We next consider the time at which the ECSF initial conditions
 should be applied. 
They strictly apply for the quantum vacuum state of an effectively massless scalar field
 ($m \ll H$) in de Sitter space. For the inflaton, $s$, we may apply the 
initial condition at any time 
 during slow rolling. So long as there is little growth of the effectively classical $s$
 fluctuation from this time until $s$ reaches $s_{c}$ (which may be 
expected as there is only the 1-loop effective potential at this 
time) the same initial perturbation may be applied at
 $s_{c}$ (or more generally up to the time at which the $U(1)_{FI}$ symmetry breaking 
field develops a classical expectation value and 
the $s$ perturbation begins to grow significantly). At $s_{c}$ the $\phi_{-}$ field 
is also massless. (However, there is a $\phi_{-}^{4}$ potential term which 
will limit the amplitude of effectively massless fluctuations to $V^{''}(\phi_{-}) \lae H^{2}$.) 
Thus a natural time to apply the equivalent classical initial conditions 
to both $s$ and $\phi_{-}$ would be at $s(t) \approx s_{c}$.

       At $s_{c}$ the mean value of the $\phi_{-}$ field is zero. 
The $U(1)_{FI}$ symmetry breaking occurs as a result of the growth of the 
$\phi_{-}$ quantum equivalent classical fluctuations 
 when the $\phi_{-}$ field acquires a negative mass squared term once $s < s_{c}$.
The amplitude of the quantum 
equivalent classical 
fluctuation is $\delta \phi_{-\;q} \approx \sqrt{2\pi} k$. 
We first note that the largest value of $k$ for $\phi_{-}$ fluctuations for 
which the $\phi_{-}$ field can be considered as 
effectively massless corresponds to $V^{''}(\delta \phi_{-\; q}) \; (\equiv 
3 g^{2} \delta \phi_{-\; q}^{2} / 2) \approx H^{2}$. Thus the spectrum of equivalent 
classical fluctuations has an upper limit, $k^{2} \lae H^{2}/3\pi g^{2}$, beyond 
which the quantum fluctuations will be suppressed by the effective mass term.  
A fluctuation can grow due to the negative mass squared term so long as its physical 
momentum satisfies
$k^{2} < |m_{\phi_{-}}^{2}|$, where $m_{\phi_{-}}^{2}  
= \left(\frac{\lambda^{2}s^{2}}{2} - g^{2} \xi \right)$. 
In addition, the mode can grow only once $|m_{\phi_{-}}^{2}| \gae H^{2}$.
Therefore, the largest amplitude mode which first begins to grow classically 
corresponds to $k^{2} \approx H^{2}$, which, for natural 
values of the $U(1)_{FI}$ gauge coupling not much smaller than 1,  
is of the same order as the upper 
limit of the spectrum of equivalent classical fluctuations. Thus we may consider 
the dominant mode to correspond to $k \approx H$ and to start growing once 
$|m_{\phi_{-}}^{2}(s)| \approx H^{2}$, defined to occur at $s = s_{1}$. 
Once the $\phi_{-}^{4}$ interaction term is included, the 
minimum of the potential in the $\phi_{-}$ direction for a given $s$ 
is at $\phi_{min} = \sqrt{2}|m_{\phi_{-}}|/g$. 
Therefore 
the minimum of the potential is at 
around the amplitude of the largest quantum equivalent 
classical fluctuation which is able to grow due to 
the negative mass squared term. Once a mode is 
given a classical value of the order of $\phi_{min}$, 
numerical solution of the classical equations of motion shows that a growing mode 
will remain close to $\phi_{min}(s)$ as $s$ decreases (Fig 1). This 
indicates that at 
any given time once $s < s_{1}$, the dominant quantum equivalent classical 
$\phi_{-}$ mode will have amplitude around $\phi_{min}(s)$. 

     This also indicates that the size of the domains of $U(1)_{FI}$ 
symmetry breaking and so seperation of the cosmic strings, 
which corresponds to the dominant $\phi_{-}$ fluctuation with $k \approx H$, 
will not be much smaller than the horizon 
($\lambda \approx 2\pi/k \approx H^{-1}$). In particular, the spacing of the cosmic strings 
will be much larger than the dynamical length
scale of the inflaton sector fields, of the order of $\xi^{-1/2}$.  
Therefore we expect that cosmic strings will be relatively rare 
and that they 
will not play a significant role in the subsequent evolution of the inflaton 
sector fields and energy density.

\begin{arrangedFigure}{1}{2}{arrangedO420}{\footnotesize{
After its initial perturbation $\phi$ tracks its minimum $\phi_{min}(s)$ as a 
function of $s$ as $s$ decreases for both values of $g$. 
(Using $\frac{\lambda}{g} = 0.14$ and
 $k=0.001s_c$.)}}
    \subFig[$g=1.0$]{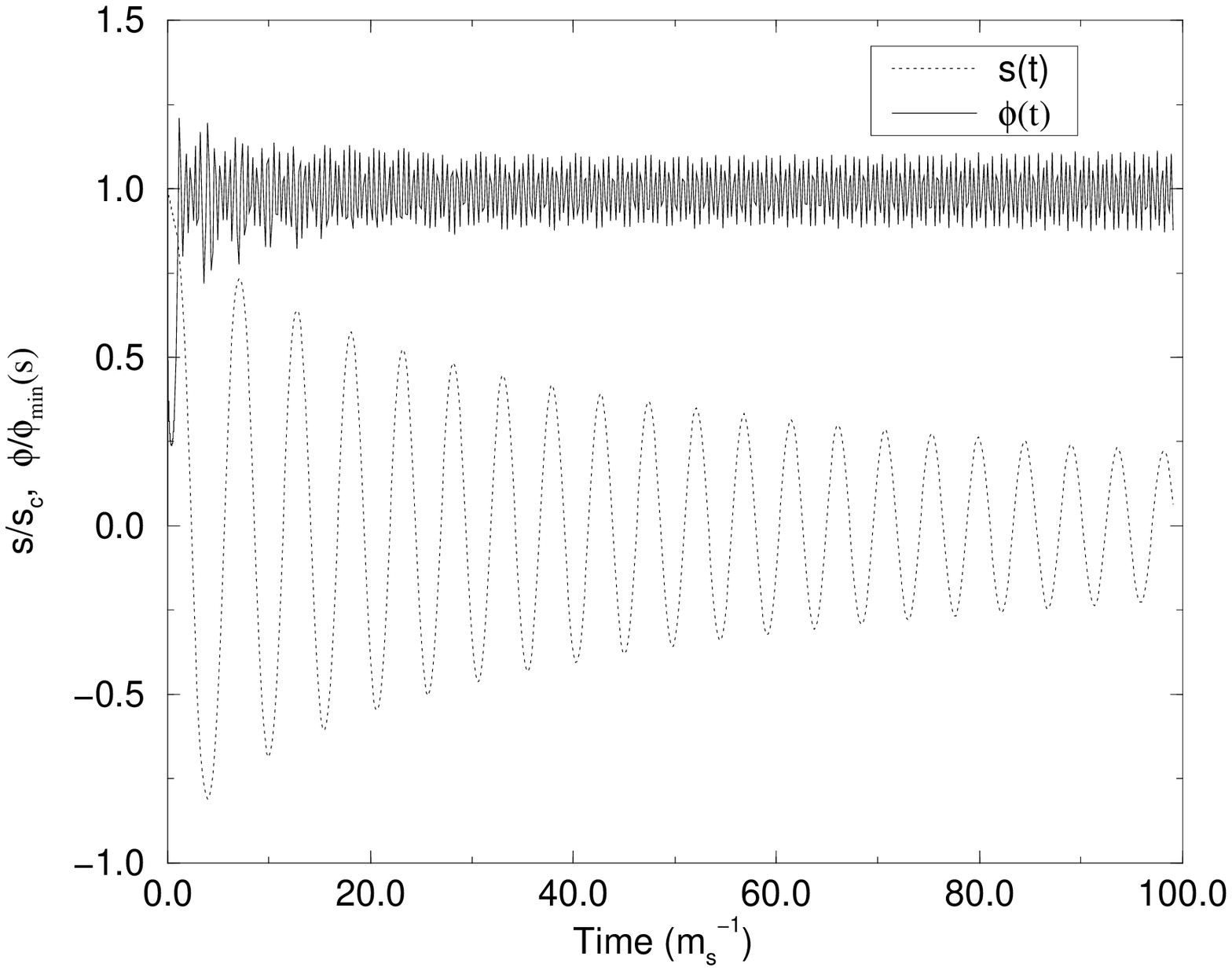}%
    \subFig[$g=0.5$]{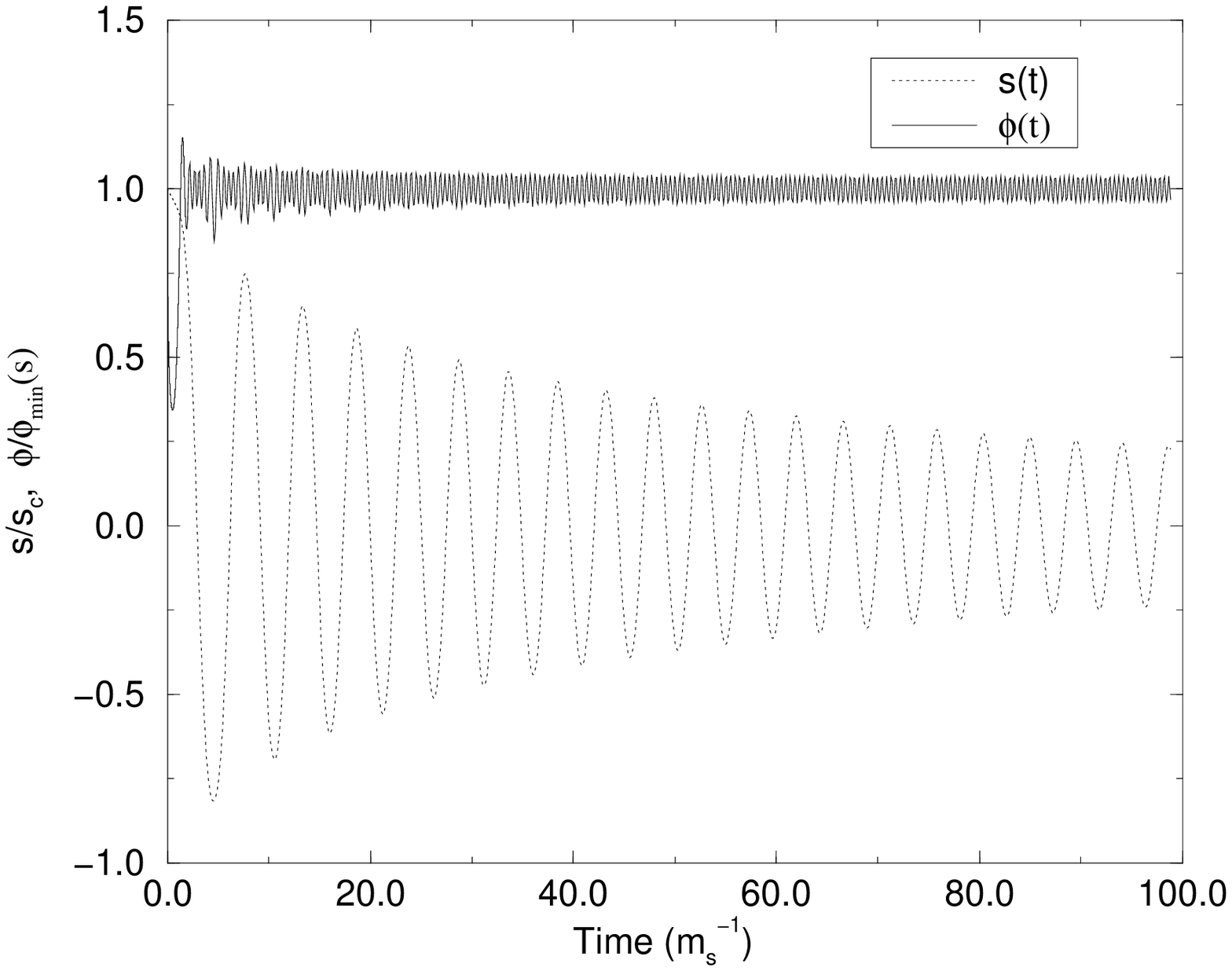}%
\setlength{\subFigureAboveCaptionSpace}{1mm} 
\end{arrangedFigure} 
\setlength{\subFigureAboveCaptionSpace}{0mm}

\subsection{Initial Conditions}

      In order to analyse the growth of $s$
perturbations in the presence of cosmic strings, it is necessary to perform a full
lattice simulation. Here
we wish to consider a simplified model without cosmic strings, which we have 
argued will not play a significant role in the growth of 
inflaton sector perturbations. This is achieved
by correlating the direction of the initial $\phi_{-}$ quantum fluctuations. Thus
we will consider an initial perturbation of the $\phi_{-}$ field with the same magnitude
 as the quantum equivalent classical fluctuations (i.e. of the order of $\phi_{min}(s)$ 
at $s_{1}$) 
but homogeneous in space. We will refer to these as 'no cosmic string' (NCS) initial 
conditions in the following.
The NCS initial conditions we consider are then 
\be{ic13} s = s_{1}  ~,\ee
\be{ic14} \delta s  = \sqrt{2\pi}k  \;\;\; ; \;\;    |\delta \dot{s}| =  
\sqrt{2\pi} k^{2} \left(\frac{H}{k}\right)  ~,\ee
\be{ic15} \phi_{-} = \phi_{min}(s_{1})   ~\ee
and
\be{ic16} \delta \phi_{-} =  0  \;\;\; ; \;\; |\delta \dot{\phi}_{-}| = 0     ~,\ee
with $s_{1}$ corresponding to $|m_{\phi}(s_{1})|^{2} \approx 
H^{2}$. 
We have set $\delta \phi_{-} = 0$ initially since the quantum equivalent 
classical $\phi_{-}$ perturbation has been rotated in space to become homogeneous. 
The initial value of the rate of rolling of the homogeneous inflaton field, 
$\dot{s}$, which plays an important role in determining the 
subsequent growth of the spatial perturbations, is obtained 
by evolving the homogeneous $s$ field from the slow-rolling regime at 
$s \gg s_{c}$ to $s_{1}$ using the 1-loop effective potential.

        The NCS initial 
conditions we have introduced are not strictly physical. Nevertheless, the
resulting growth of the inflaton perturbation has a 
general physical interpretation, namely it is the maximum
 possible growth. This follows since the classical 
growth of inflaton perturbations is due to
 the non-zero $\phi_{-}$ expectation value as a function of $s$; in the limit $\phi_{-}
 = 0$ there
 is no significant growth since $s$ then only experiences a 1-loop effective 
potential as in the case $s > s_{c}$. The effect of the cosmic 
strings is to reduce 
the value of $\phi_{-}$ at a given point in space relative to the case with no cosmic
 strings (the $U(1)_{FI}$ symmetry breaking field vanishing at the centre of the strings), 
so reducing the degree of $U(1)_{FI}$ symmetry breaking on average. Therefore we expect the
 growth of inflaton perturbations to be suppressed relative to the case 
with no cosmic strings. For the particular case of D-term inflation 
we have also argued that cosmic strings should be 
relatively rare on sub-horizon scale and so should not play a major role in 
determining the evolution of the energy density in the bulk between the strings. 
In this case the NCS initial conditions should be a 
good approximation to the fully realisitic case.

\section{Semi-Analytical Model for Spatial Perturbation Growth}

    In this section we consider the growth of 
spatial perturbations using an analytical ansatz 
for the spatial dependence of the perturbation modes. 
Our method will be to study the evolution of a single perturbation mode 
of the 
$s$ and $\phi_{-}$ field, using an ansatz for the functional form of the space 
dependence of the fields. We first discuss the ansatz for a generic real scalar field $\Phi$. 
The field we consider is assumed to initially have the form $\Phi(\vec{x},t) = \Phi(t) 
+ \delta \Phi(t) \sin (\vec{k}.\vec{x})$, where $\Phi(t)$ is the homogeneous field. 
In general, we would expect the spatial position of the 
extrema of $\Phi(\vec{x},t)$ at later times to remain the same as their initial positions. 
Thus we expect that $\Phi(\vec{x},t) = \Phi(t) + f(t)g(\vec{x})$, where $g(\vec{x})$ has 
the same periodicity in space as the initial perturbation. If the 
growth of the perturbation was strictly linear, then
the correct choice would be $g(\vec{x}) = \sin(\vec{k}.\vec{x})$.
However, it is possible that non-linearity could play a role in the growth of the perturbation, 
in which case this assumption must be considered an ansatz, whose validity depends on how well the 
true interpolating function $g(\vec{x})$ between the maxima and minima of $\Phi(\vec{x},t)$ in space can 
 be modelled by $\sin(\vec{k}.\vec{x})$. The match need not be exact for the physics of
perturbation growth to be reasonably well modelled. 

    The equation of motion of $\Phi$ is 
\be{sa1} \ddot{\Phi} + 3 H \dot{\Phi} - \frac{\vec{\nabla}^{2}}{a^{2}} \Phi 
= - \frac{\partial V(\Phi)}{\partial \Phi}    ~.\ee
We introduce the ansatz $\Phi(\vec{x},t) = \Phi(t) 
+ \delta \Phi(t) \sin(\vec{k}.\vec{x})$, 
in which case the equations of motion become
\be{sa2} \ddot{\Phi} + 3 H \dot{\Phi} 
+ \frac{\vec{k}^{2}}{a^{2}} \delta \Phi \sin(\vec{k}.\vec{x})= 
- \frac{\partial V\left(\Phi\right) }{\partial \Phi}   ~.\ee
Let $\Phi_{+}$ and $\Phi_{-}$ 
be values of $\Phi(\vec{x},t)$ at points 
in space ($\sin(\vec{k}.\vec{x}) = \pm 1$)
corresponding initially to the 
maximum and minimum of $\Phi(\vec{x},t)$.  
Thus $\Phi_{+} = \Phi(t) + 
\delta \Phi(t)$ and $\Phi_{-} = \Phi(t) - \delta \Phi(t)$. The equations at these 
points in space are then 
\be{sa3} \ddot{\Phi}_{+} + 3 H \dot{\Phi}_{+} 
+ \frac{\vec{k}^{2}}{a^{2}} \delta \Phi = 
- \left. \frac{\partial V\left(\Phi\right) }{\partial \Phi} \right|_{\Phi_{+}}   ~\ee 
and
\be{sa4} \ddot{\Phi}_{-} + 3 H \dot{\Phi}_{-} 
- \frac{\vec{k}^{2}}{a^{2}} \delta \Phi = 
\left. - \frac{\partial V\left(\Phi\right) }{\partial \Phi} \right|_{\Phi_{-}}   ~.\ee 
Thus with $\delta \Phi = (\Phi_{+} - \Phi_{-})/2$ the equations for the evolution of the 
field at points in space corresponding to maxima and minima of $\Phi$ are
\be{sa3} \ddot{\Phi}_{+} + 3 H \dot{\Phi}_{+} 
+ \frac{\vec{k}^{2}}{2 a^{2}} (\Phi_{+} - \Phi_{-})  = 
\left. - \frac{\partial V\left(\Phi\right) }{\partial \Phi}\right|_{\Phi_{+}}   ~\ee
and 
\be{sa4} \ddot{\Phi}_{-} + 3 H \dot{\Phi}_{-} 
- \frac{\vec{k}^{2}}{2 a^{2}} (\Phi_{+} - \Phi_{-}) = 
\left. - \frac{\partial V\left(\Phi\right) }{\partial \Phi}\right|_{\Phi_{-}}   ~.\ee 
Thus we have only to solve two coupled equations which are purely functions of $t$. 
We will refer to this as the semi-analytical model for perturbation growth.

    We can generalise the semi-analytical model to the case of hybrid inflation with 
two scalar fields. We consider an ansatz of the form 
$s(\vec{x},t) = s(t) + \delta s(t) \sin(\vec{k}.\vec{x})$ 
and $\phi(\vec{x},t) = 
\phi(t) + \delta \phi(t) \sin(\vec{k}.\vec{x})$. 
At $s(t) = s_{1}$ the $\phi$ 
field at a point in space corresponding to the minimum of 
$s(\vec{x},t)$, $s_{-}$, will experience the largest 
symmetry breaking negative mass squared term. 
Therefore at this point $\phi(\vec{x},t)$ begins to 
grow first, corresponding to $\phi_{+}$. (In accordance with NCS boundary conditions, 
we assume the symmetry breaking direction is correlated.)  
We therefore denote the fields at this point in space by $(s_{-}, \phi_{+})$, 
whilst the fields at the maximum of $s(\vec{x},t)$ are $(s_{+},\phi_{-})$. 
(The spatial dependence of the $s$ perturbation is therefore
imprinted on the $\phi$ field.)
This is illustrated in Figure 2. The semi-analytic equations are then
\be{gsp4} \ddot{s}_{+} + 3H\dot{s}_{+} + \frac{k^2}{2 a^2}(s_{+}-s_{-}) =
\left. - \frac{\partial V(s,\phi)}{ \partial s} \right|_{(s_{+},\phi_{-})}
~,\ee
\be{gsp5}
\ddot{\phi}_{-} + 3H\dot{\phi}_{-} - \frac{k^2}{2a^2}(\phi_{+}-\phi_{-}) =
\left. - \frac{\partial V(s,\phi)}{\partial \phi} \right|_{(s_{+},\phi_{-})}
~,\ee
and
\be{gsp6}
\ddot{s}_{-} +3H\dot{s}_{-} - \frac{k^2}{2a^2}(s_{+}-s_{-}) =
\left. - \frac{\partial V(s,\phi)}{\partial s} \right|_{(s_{-},\phi_{+})}
~,\ee
\be{gsp7}
	\ddot{\phi}_{+} + 3H\dot{\phi}_{+} 
+ \frac{k^2}{2a^2}(\phi_{+}-\phi_{-})=
\left. - \frac{\partial V(s,\phi)}{\partial \phi} \right|_{(s_{-},\phi_{+})}
~.\ee
Then $\delta s = (s_{+}-s_{-})/2$ and $\delta \phi = (\phi_{+}-\phi_{-})/2$, whilst the 
homogeneous mean fields are given by $s(t) = (s_{+}+s_{-})/2$ and  
$\phi(t) = (\phi_{+}+\phi_{-})/2$. \\

\begin{figure}[htbp]
\begin{center}
\includegraphics[width=0.8\textwidth]{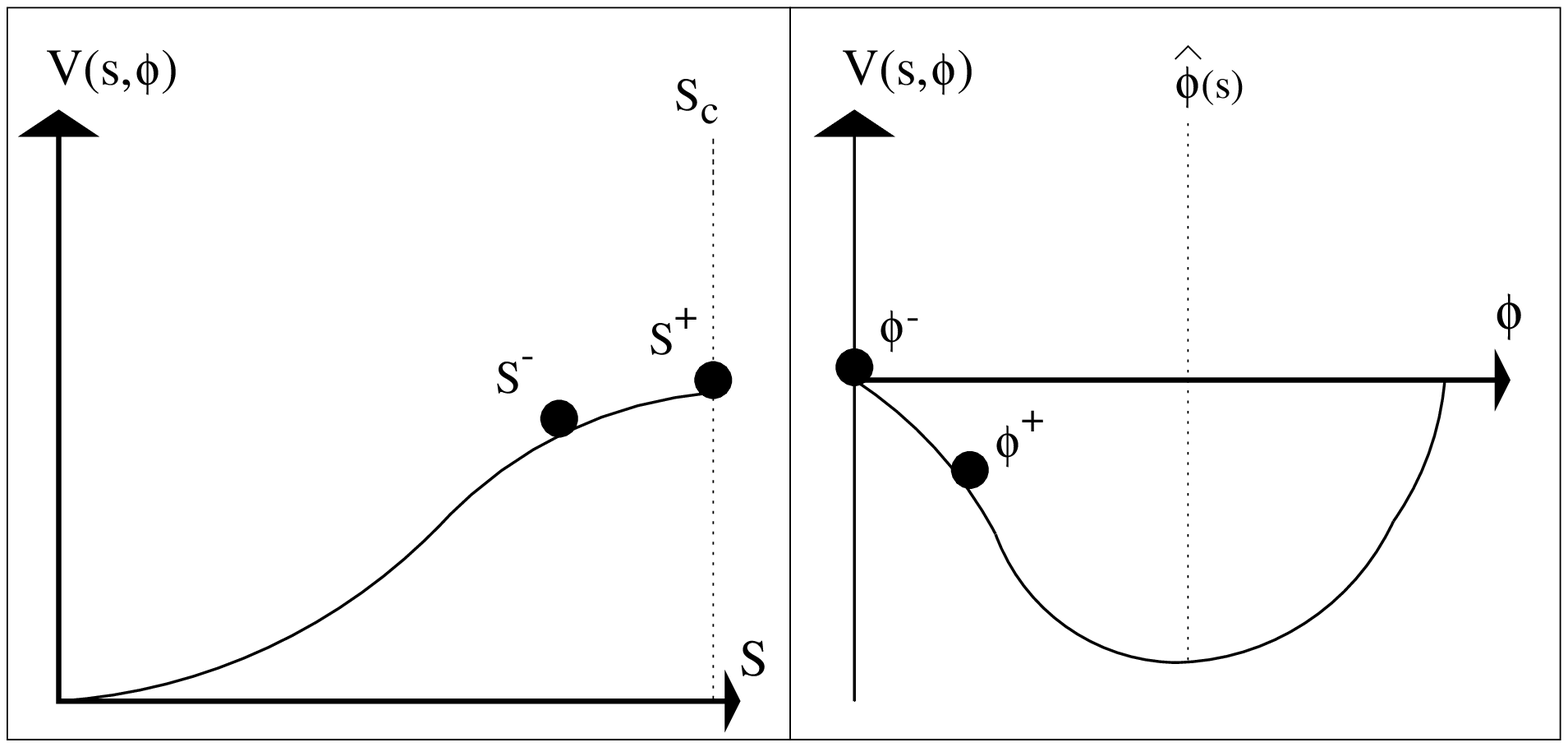}
\caption{\footnotesize{Schematic illustration of the time evolution of $\phi$ and
$s$ as the inflaton $s$ passes from slow-roll 
phase to $U(1)_{FI}$ symmetry breaking phase.}}
\label{fig:fig2}
\end{center}
\end{figure}

\section{Results for Perturbation Growth}

	We next investigate the growth in inflaton perturbations 
using the NCS initial conditions and the semi-analytical
ansatz for the evolution of the spatial perturbations. We will focus 
on the case $g=1$ but include $g=0.5$ for comparison.

      In the figures we show the evolution of the homogeneous mean field $s(t)$ and the 
amplitude of spatial perturbation $\delta s(t) = (s_{+} - s_{-})/2$ 
for the semi-analytical model using the D-term hybrid inflation scalar
potential with 1-loop
corrections. (For clarity we show  
the magnitude of $\delta s(t)$ in the figures, such that $\delta s(t)$ is 
always positive.) 
The homogeneous inflaton field is evolved from the  
slow-rolling regime with initially $s \approx 20 s_{c}$ 
whilst the NCS boundary conditions for the perturbations are introduced at $s_{1}$. 
We expect that the growth of the spatial perturbations will become fully non-linear 
with the formation of non-topological solitons (`fragmentation`) once the spatial 
perturbation amplitude is similar to the homogeneous mean field, $\delta s(t)/s(t) \approx 1$. 
 
        In Figure 3 and 4 we show the evolution of $\delta s/s(t)$ for a mode 
of fixed wavenumber $k = 0.01 s_{c}$ as $\lambda/g$ is increased for the case $g=1$. 
(We consider $a =1$ at $s = s_{c}$, such that $k$ is the physical wavenumber at this time.)
In Figure 3a) with $\lambda/g =0.1$ there is little growth of $\delta s(t)$ whilst the mean field amplitude 
$s(t)$ decreases due to expansion. In Figure 3b) with $\lambda/g = 0.12$ significant growth of the spatial 
perturbation begins to be seen. In Figure 4a) with $\lambda/g = 0.14$ 
we see that the spatial perturbation 
amplitude increases rapidly to become equal to the mean field, $\delta s(t)/s(t) \approx 
1$, in approximately one oscillation cycle of the homogeneous field. Finally, in 
Figure 4b) with $\lambda/g = 0.2$ the spatial perturbation grows more rapidly and dominates the
 homogeneous field at all times after about one 
oscillation cycle of the homogeneous field. 

     These results indicate that the 
the growth of quantum fluctuations of the inflaton sector fields to form
 non-topological solitons can be rapid in realistic D-term inflation models, occuring in 
not much more than the time for a single coherent oscillation of the 
inflaton field. 
This is consistent with a tachyonic preheating picture of perturbation 
growth \cite{tp} rather than an inflaton
 condensate fragmentation interpretation \cite{icf}.

\begin{arrangedFigure}{1}{2}{arrangedO521}{\footnotesize{$k=0.01s_c$. As we increase 
 $\frac{\lambda}{g}$ from 0.10 to 0.12 we start to see growth in
 the perurbation amplitude though it is still well within the linear regime.}}
    \subFig[$\frac{\lambda}{g} = 0.10$]{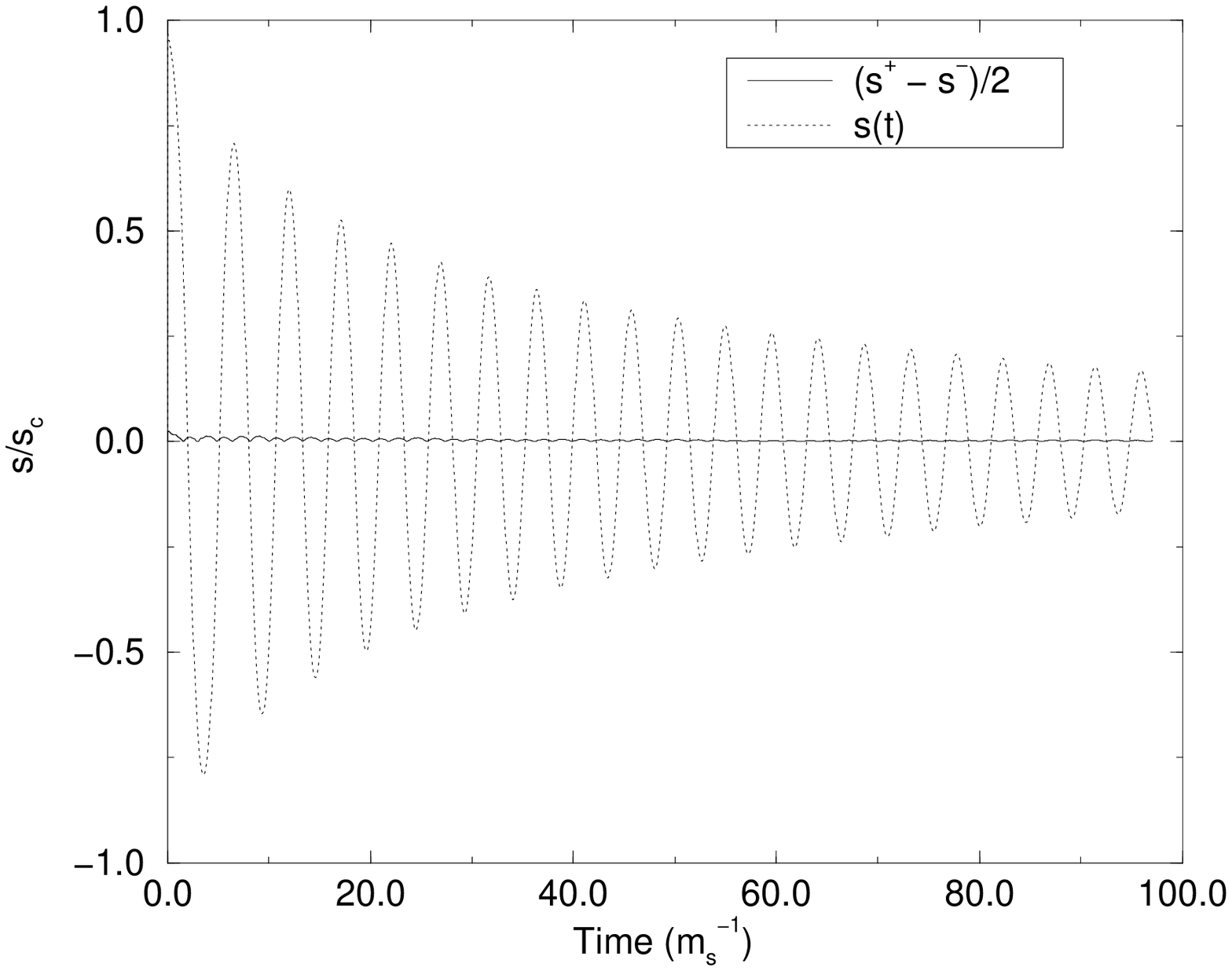}%
    \subFig[$\frac{\lambda}{g} = 0.12$]{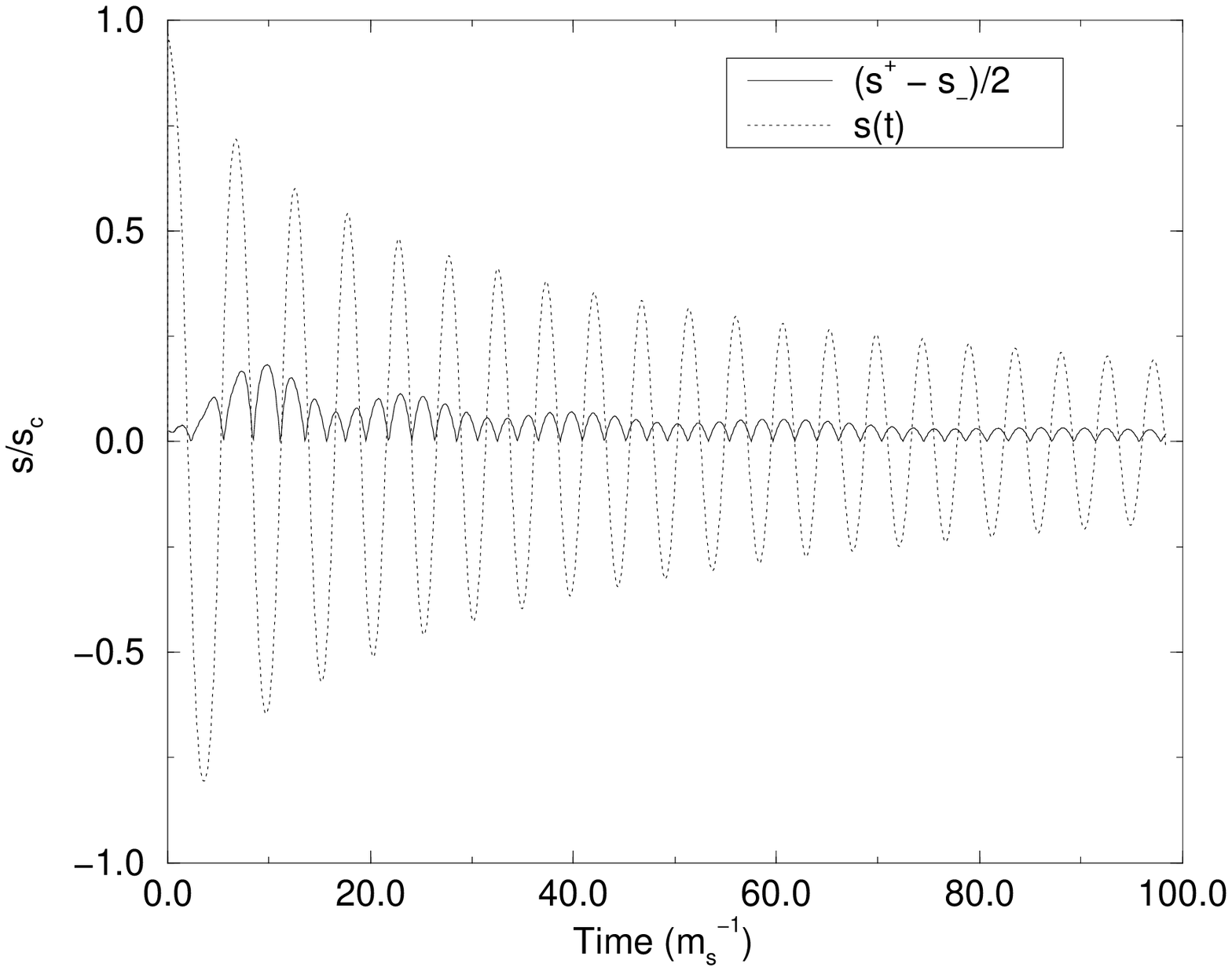}%
\newSubFig{arrangedO522}{\footnotesize{$k=0.01s_c$. (a) shows the critical value of 
$\frac{\lambda}{g}$ for which the perturbation just becomes fully non-linear. Any further increase
((b)) pushes the perturbation deeper into the non-linear regime.}}
    \subFig[$\frac{\lambda}{g} = 0.14$]{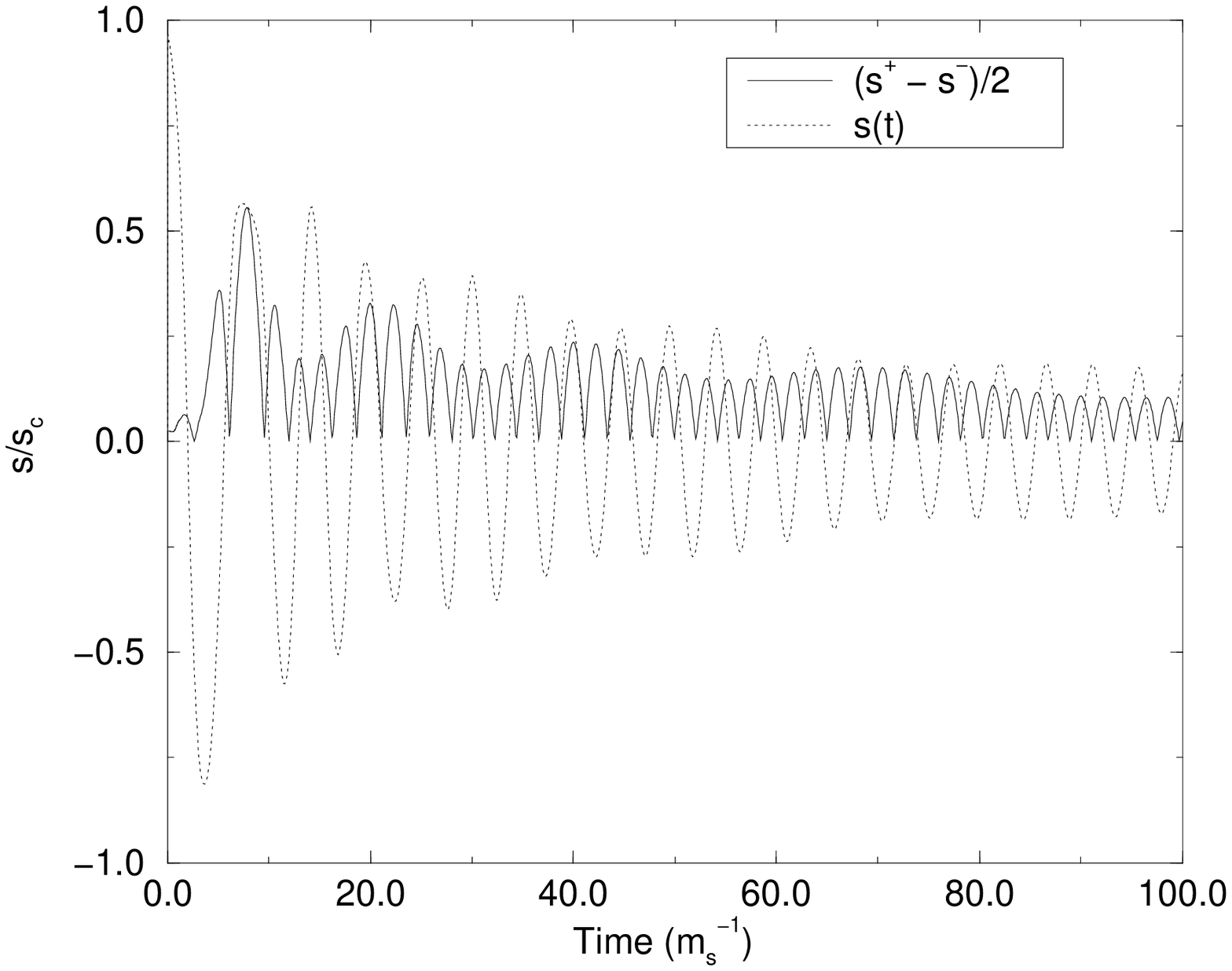}%
    \subFig[$\frac{\lambda}{g} = 0.2$]{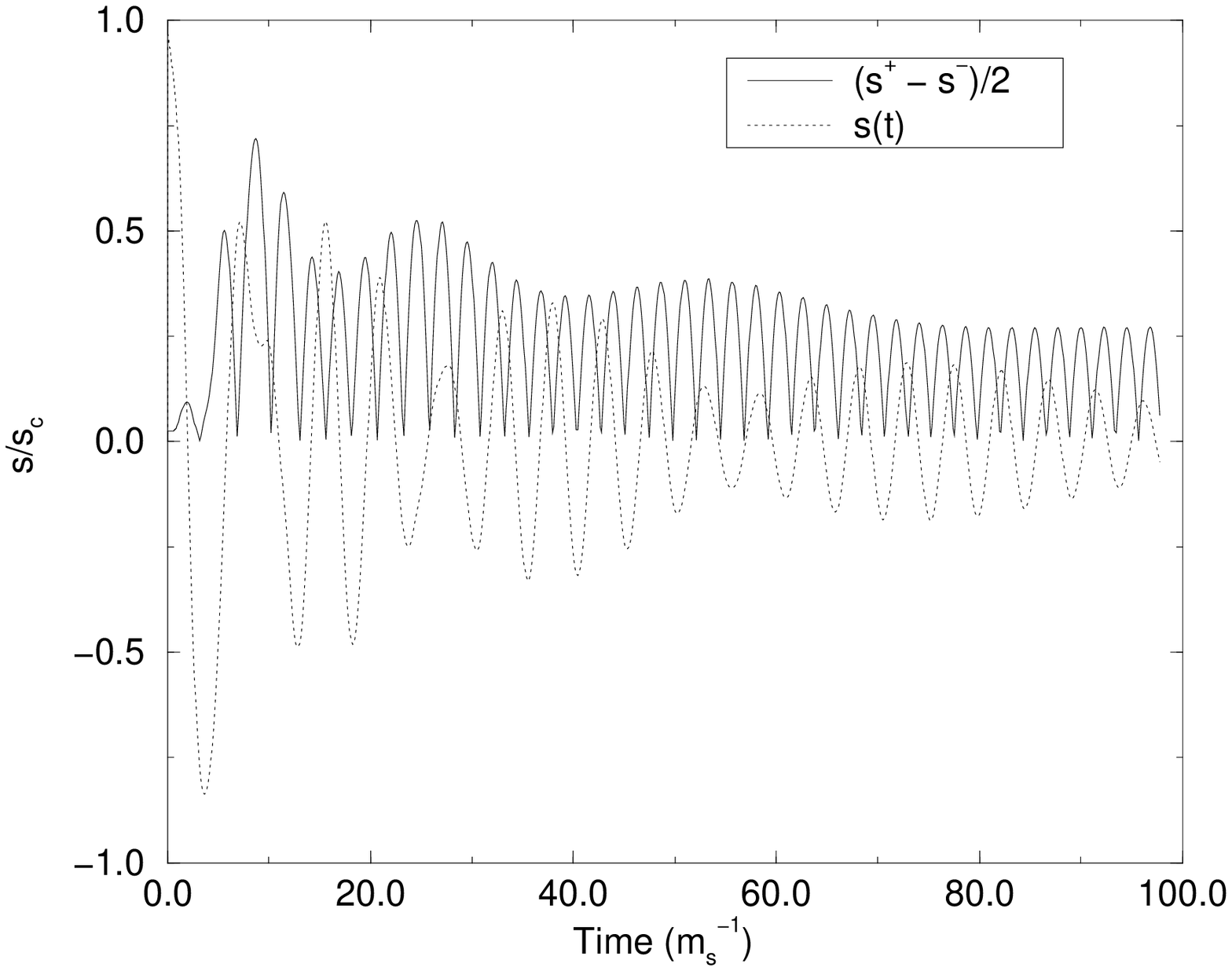}%
\setlength{\subFigureAboveCaptionSpace}{7mm} 
\end{arrangedFigure} 
\setlength{\subFigureAboveCaptionSpace}{0mm} 

        We next consider the case of  fixed $\lambda/g = 0.14$ and investigate the
effect of the variation of the mode $k$ on the growth of the perturbations. 
Figure 5 shows the transition from a) the case of almost no-growth at $k = 0.001s_{c}$ to
b) fully non-linear growth ($\delta s / s(t) \gae 1.0$) 
at $k \approx 0.005s_c$. 
In Figure 6 from a) to b) a dramatic damping in the growth of the perturbation is
seen as $k$ increases above $0.01 s_{c}$.  
Since the effect of large
gradient terms in the equations of motion is to damp the growth of the
perturbation, this would indicate that 
we have passed the maximum value, $k_{max}$,
for which growth of inflaton perturbations is possible. The dominant mode for a 
given $\lambda/g$ will correspond to that which grows to become 
fully non-linear first, which will correspond to $k \approx k_{max}$. 

In Figures 7 and 8 we consider $g = 0.5$ whilst keeping $\lambda/g = 0.14$. 
Again we vary $k$ to show the transition from no growth to fully non-linear growth. 
In Figure 7b) we find that for $k = 0.001s_{c}$ the mode can grow to become
 non-linear but in a time much longer than the period of coherent oscillations. Growth
 of this mode would be well described by the inflaton condensate fragmentation approach \cite{icf}. 
However, it is the mode which grows most rapidly that will reach non-linearity first and so
 determine the length scale of the condensate lumps. Typically this
 dominant mode reaches non-linearity in a few oscillations.  
    
      Comparing Figure 8a) with Figure 6a), we see that exactly the same maximum 
growth of the perturbation is obtained at $k$ close to $k_{max}$ for both $g = 0.5$ 
and $g = 1.0$ when $\lambda/g$ is fixed. This shows that the maximum growth of the
 perturbations is dependent only upon the ratio $\lambda/g$. We also see that 
$k_{max} \approx 0.007 s_{c}$ for $g = 0.5$, compared with 
$k_{max} \approx 0.015 s_{c}$ for $g = 1$, 
indicating that $k_{max} \propto g$ for fixed $\lambda/g$.  

   An important quantity is the lower limit on $\lambda/g$ for which the 
perturbations can grow sufficiently to achieve fragmentation. We find that for 
$\lambda/g \lae 0.09$ fragmentation does not occur for any value of $k$. 
Thus for $\lambda/g \gae 0.09$ we expect fragmentation into non-topological solitons to 
occur in D-term inflation. The value of $k_{max}$ corresponding to the limiting case 
$\lambda/g \approx 0.09$ is $k_{max} \approx 0.008s_{c}$. 

       Although the above results strictly apply to the case of D-term inflation,
 we note that in the limit where the classical potential becomes identical to F-term
 inflation, $\lambda = \sqrt{2}g$, condensate 
fragmentation is certain to occur in D-term inflation. 
This strongly suggests that fragmentation will be an inevitable 
feature of F-term inflation models. \\

         It is interesting to compare these results with the analytical results obtained in the 
inflaton condensate fragmentation approach \cite{icf}. Although the approximations 
made in \cite{icf} are not valid for perturbations which grow rapidly on the time scale of the coherent
oscillations, we nevertheless find that the dependence of $k_{max}$ and the 
fragmentation condition on $\lambda$ and $g$ are well described by the analytical expressions in 
\cite{icf}. $k_{max}$ is given by  
\be{ic1} \frac{ \sqrt{3} \lambda^{2}}{2g} s_{c}   ~,\ee 
where we have used $R = s_{c}$ and $a_{o} = a$ in Equation 29 of \cite{icf}, consistent with $k_{max}$ 
at the onset of coherent oscillations. 
Thus we find $k_{max} \propto g$ for fixed $\lambda/g$. For $\lambda/g = 0.14$ 
we find from \eq{ic1} that $k_{max} = 0.008 s_{c}$($0.016 s_{c}$) for $g= 0.5$(1.0), in reasonable 
agreement with our numerical results. The 
condition for fragmentation to occur from the inflaton condensate fragmentation analysis is \cite{icf}
\be{ic2} \frac{\lambda}{g} \gae \frac{4 \sqrt{2} \xi^{1/2} \beta}{M} \approx 0.2 
\left( \frac{\beta}{10} \right)    ~,\ee
where $\beta = \log(s_{c}/\delta s(t_{o}))$. This depends only on the ratio 
$\lambda/g$, in agreement with the numerical analysis. (The actual value of the
 lower bound from \eq{ic2} can only be taken as suggestive, as the various 
approximations made introduce 'factor of 2' 
uncertainties.) We also note that the rate 
of growth of perturbation from the inflaton condensate fragmentation analysis 
(given by $\alpha$ (Equation 
27) in \cite{icf}) is proportional to $k$ for values up to $k_{max}$, 
beyond which there is no perturbation growth, in qualitative agreement 
with the numerical results. 

    The broad agreement of the numerical $k_{max}$ with the inflaton condensate 
fragmentation expression suggests that
 the exponential terms governing the growth of perturbations have the same 
parameter dependence in the realistic case 
as in the inflation condensate fragmentation analysis, although 
there is no reason to expect the perturbation 
growth at $k \approx k_{max}$ to be quantitatively well-described  
by this analysis. 

        Finally, we would like to comment on the possible consequences of including the
 $\Phi_{+}$ field in the analysis. During inflation 
$m_{\Phi_{+}}\geq \sqrt{2} g\xi^{1/2} \gg H$. Therefore we expect that 
the homogeneous field will satisfy $\Phi_{+} = 0$ at the end of inflation. In this case
 the homogeneous $\Phi_{+}$ field will not evolve since $\Phi_{+} = 0$  is an
 extremum of the potential for all values of $S$ and $\Phi_{-}$. A quantum fluctuation
 of $\Phi_{+}$ could grow if the mass squared of $\Phi_{+}$ at $\Phi_{+} = 0$ 
were to become negative during the post-inflation evolution of the $S$ and $\Phi_{-}$
 fields. This could briefly occur as $|S|$ approaches zero if the value
 of the oscillating $\Phi_{-}$ field was also sufficiently larger than its minimum as a
 function of $|S|$ at this time.  
 However, the initial quantum de Sitter fluctuation of $\Phi_{+}$ will be suppressed by
 its large mass compared with $H$ during inflation. In addition, the brief growth
 as $|S|$ passes through zero will result in only a small growth of the quantum fluctuation of $\Phi_{+}$.
This is particularly true
 since the dominant $S$ modes reach fragmentation within only a few oscillations of
 the homogeneous $S$  field.  
Therefore we expect that the $\Phi_{+}$ field will have a negligible effect on the
 evolution of the energy density during the post-inflation era and that we may consider
 $\Phi_{+} = 0$ throughout.

 \begin{arrangedFigure}{1}{2}{arrangedO523}{\footnotesize{$\frac{\lambda}{g} = 0.14$, $g=1$.
 As expected, for a very small value of $k$ ((a)) there is no growth in the perturbation as the
gradient term in the equations of motion are negligable. As $k$ is increased ((b)) we get the expected
evolution into the non-linear regime.}}
    \subFig[$k=0.001s_c$]{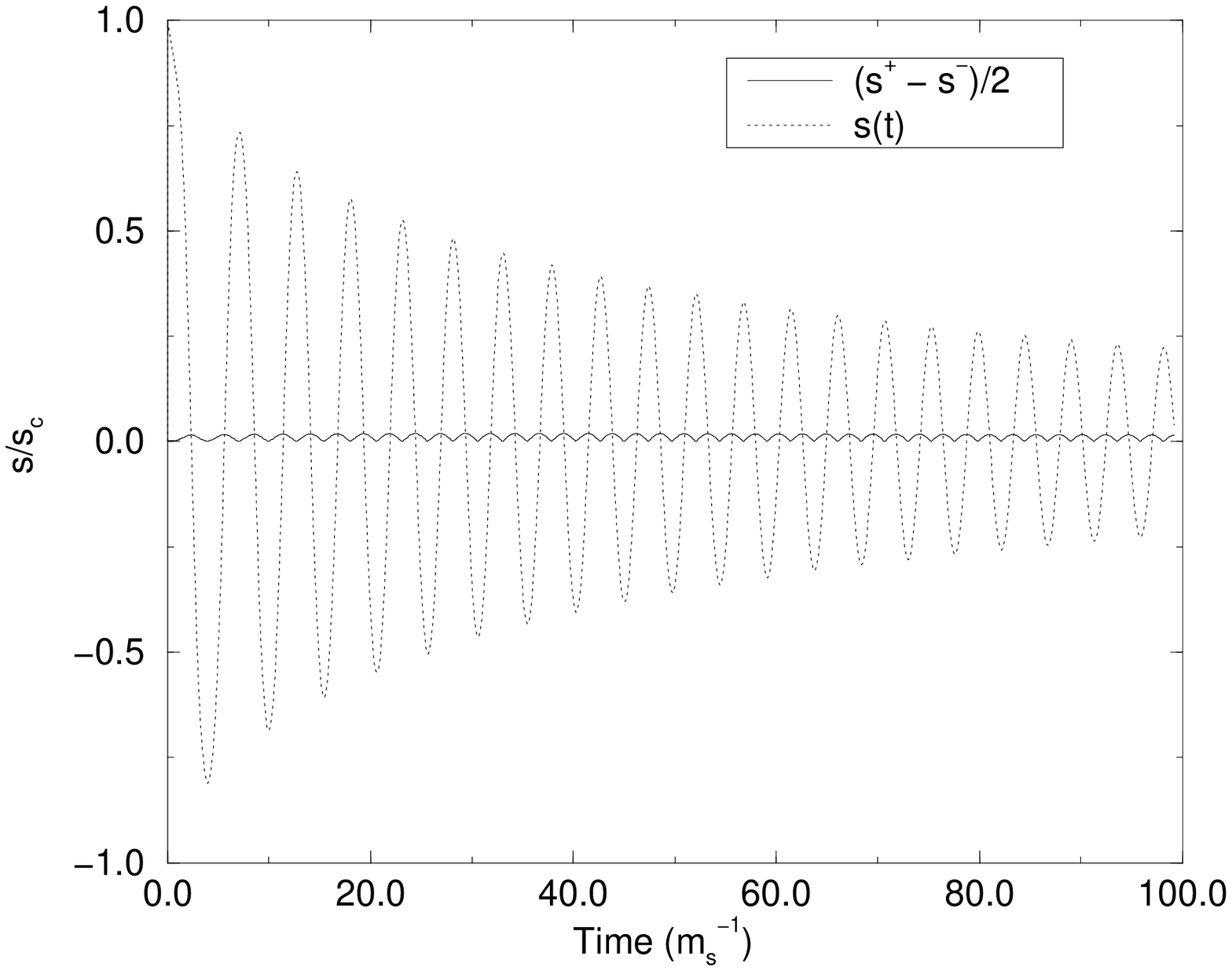}%
    \subFig[$k=0.005s_c$]{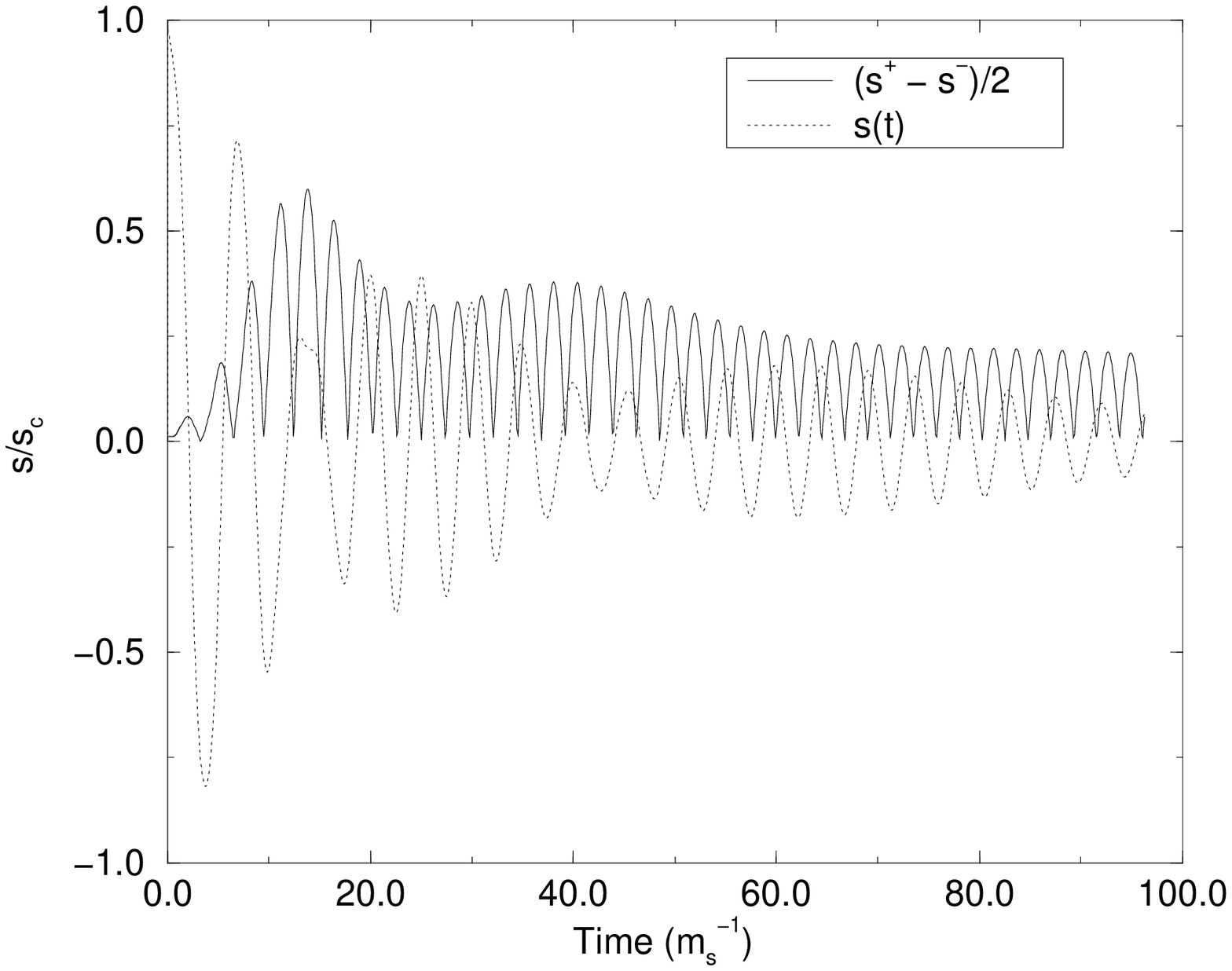}%
\newSubFig{arrangedO524}{\footnotesize{$\frac{\lambda}{g} = 0.14$, $g=1$. Further increase ((a))
 gives us the maximum value of $k$ for which full non-linearity occurs earliest at this value of
$\frac{\lambda}{g}$. This corresponds to the dominant mode leading to fragmentation. 
Finally, (b) illustrates the case $k \approx k_{max}$ 
with suppressed perturbation growth.}}
    \subFig[$k=0.010s_c$]{0.14.eps}%
    \subFig[$k=0.015s_c$]{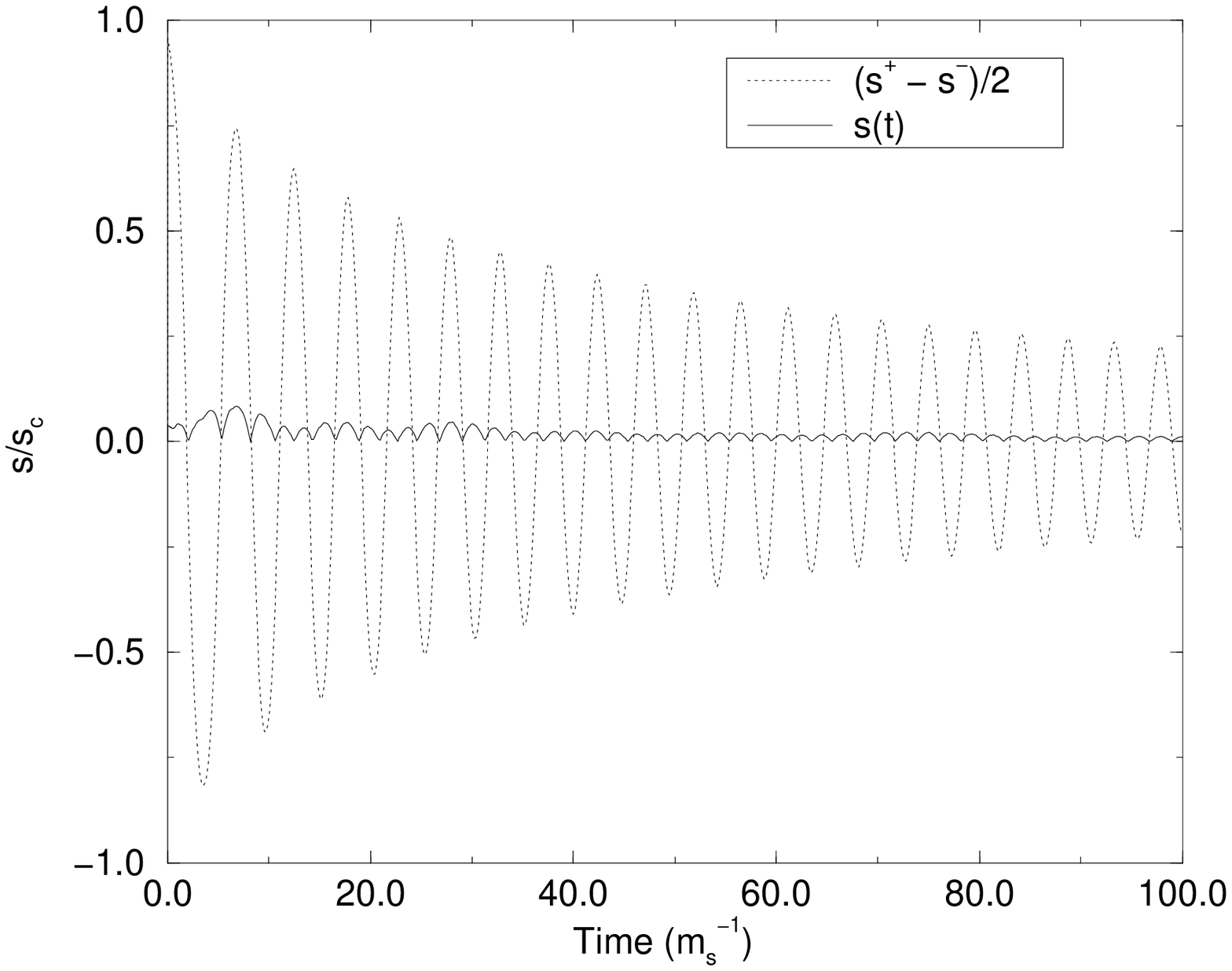}%
\setlength{\subFigureAboveCaptionSpace}{7mm} 
\end{arrangedFigure} 
\setlength{\subFigureAboveCaptionSpace}{0mm} 

\begin{arrangedFigure}{1}{2}{arrangedO524}{\footnotesize{$\frac{\lambda}{g} = 0.14$, $g=0.5$.
 As before we see the same increase in the rate of perturbation growth as we increase the
initial mode size $k$. (b) shows a mode with slow growth towards full non-linearity.}}
    \subFig[$k=0.0005s_c$]{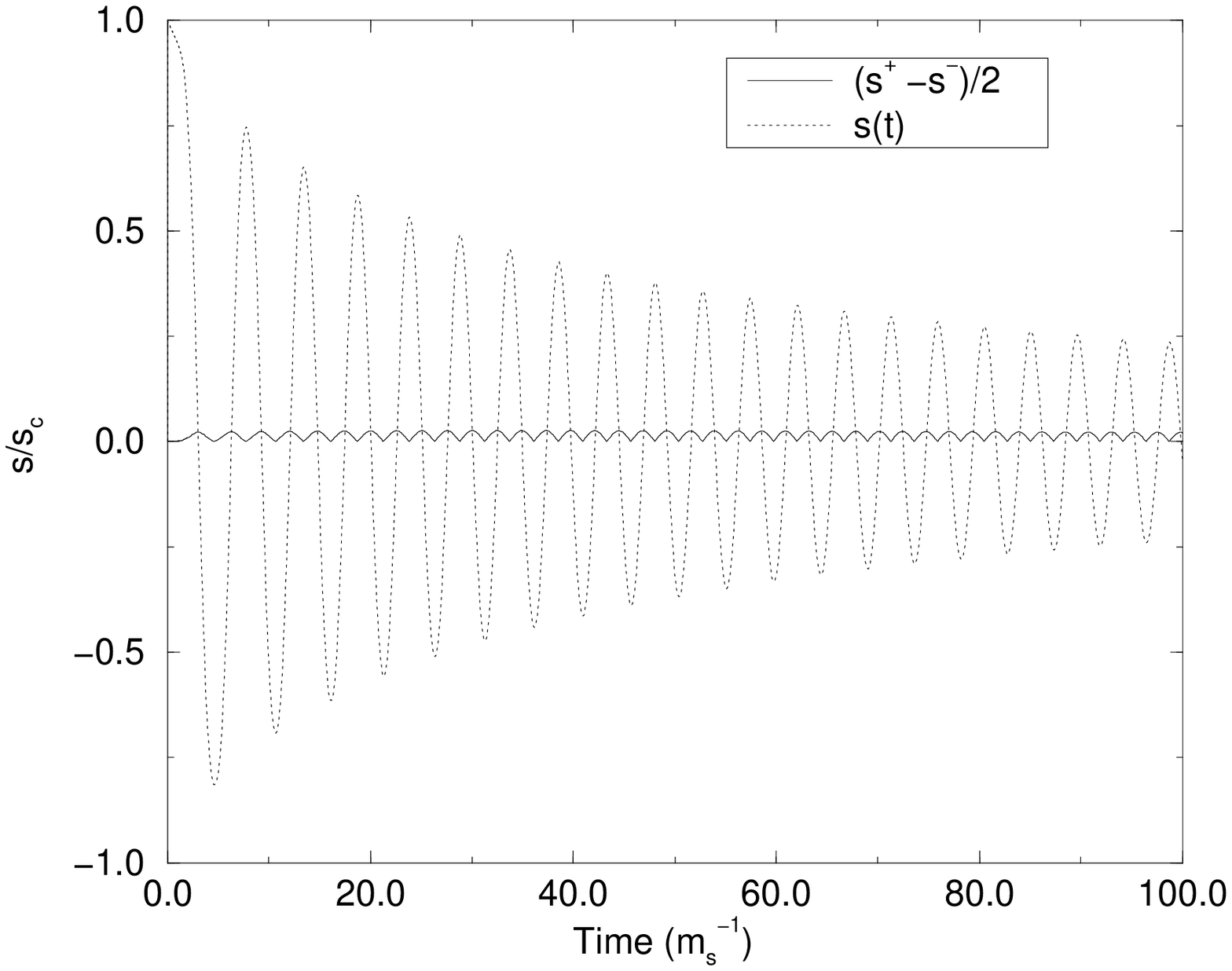}%
    \subFig[$k=0.001s_c$]{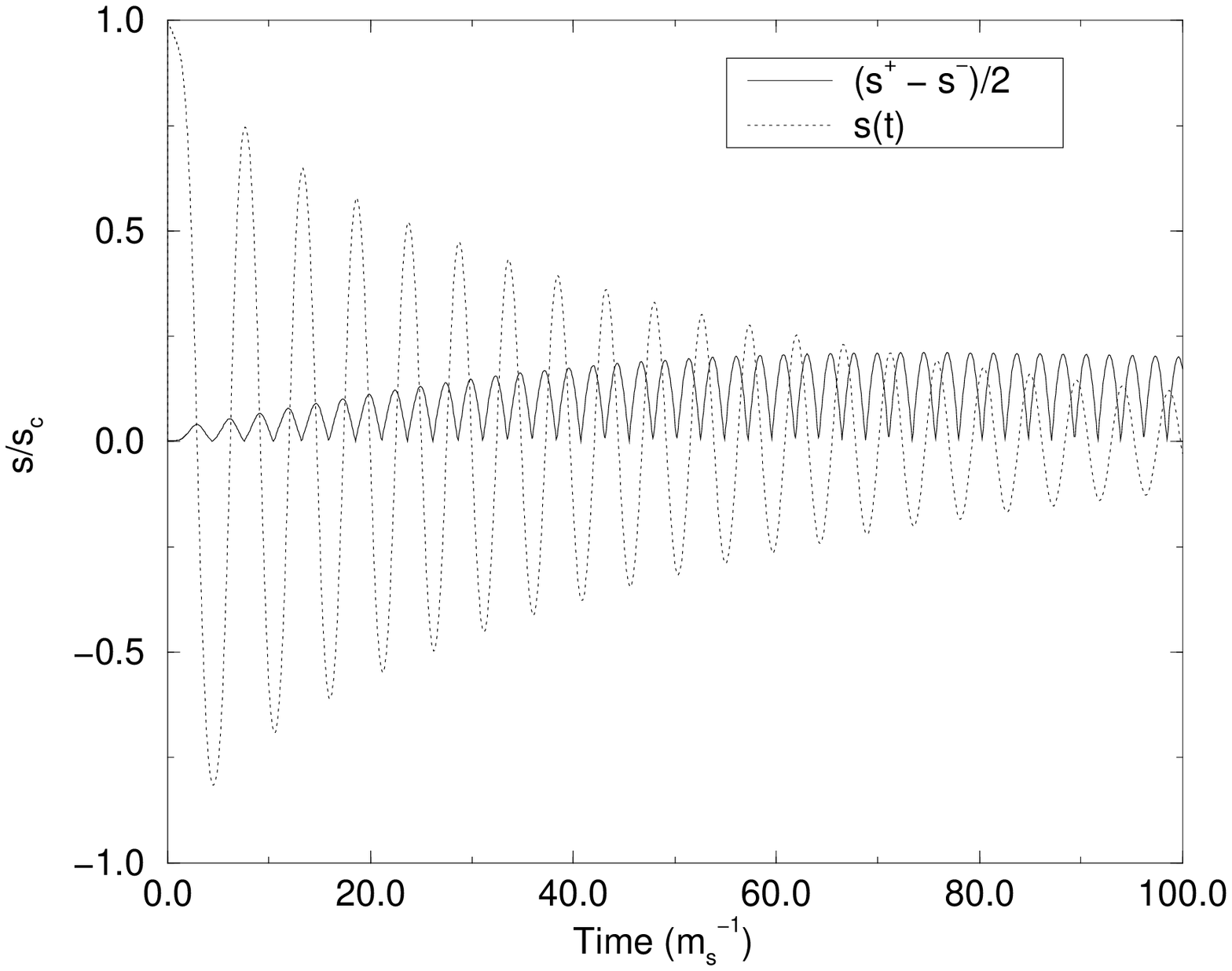}%
\newSubFig{arrangedO524}{\footnotesize{$\frac{\lambda}{g} = 0.14$, $g=0.5$. (a) shows the
limiting case with $k \approx k_{max}$ where the perturbation just reaches full non-linearity. 
Further slight increase in
 the mode size to $k \geq k_{max}$ ((b)) suppresses growth in the perturbation as before.}}
    \subFig[$k=0.005s_c$]{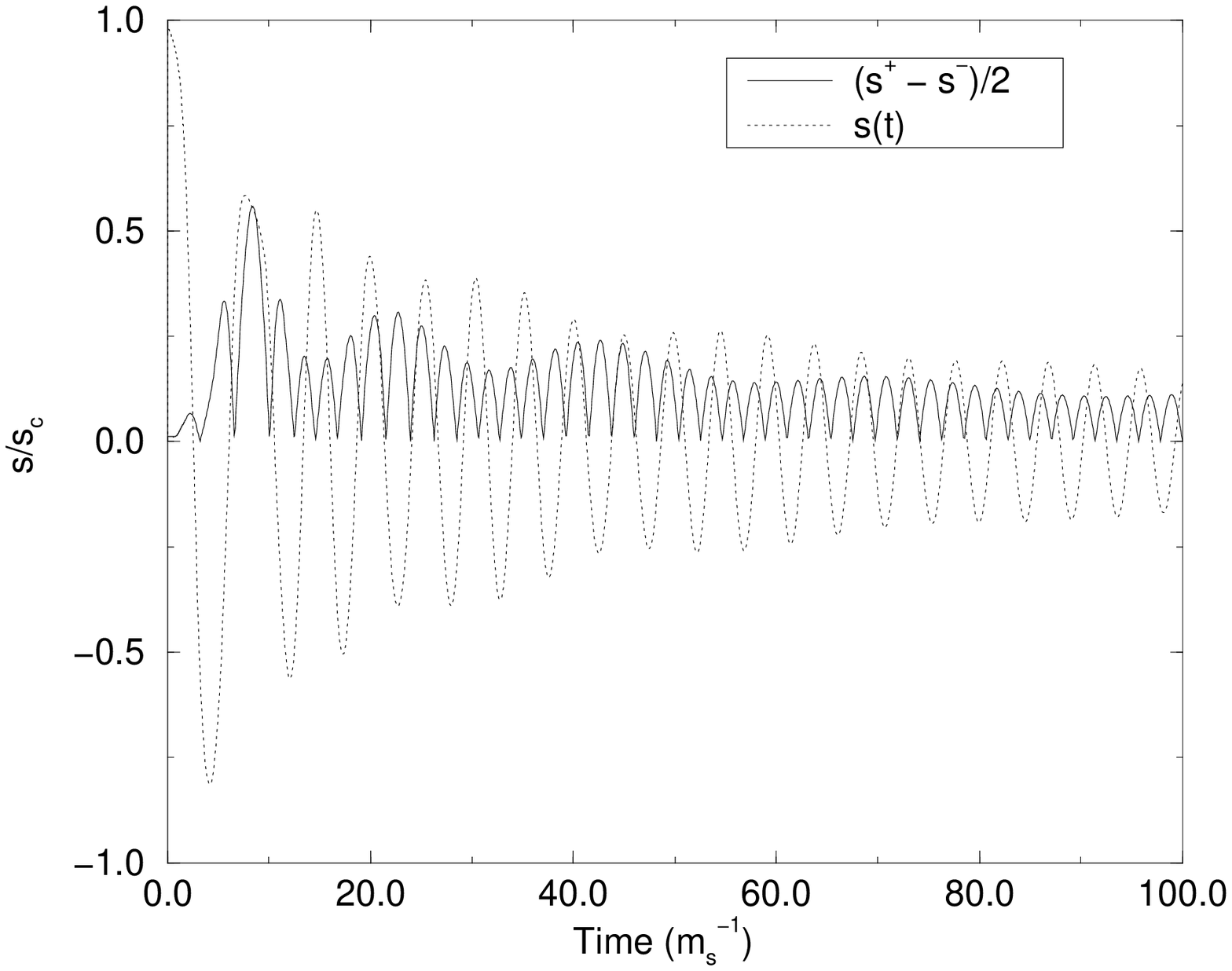}%
    \subFig[$k=0.007s_c$]{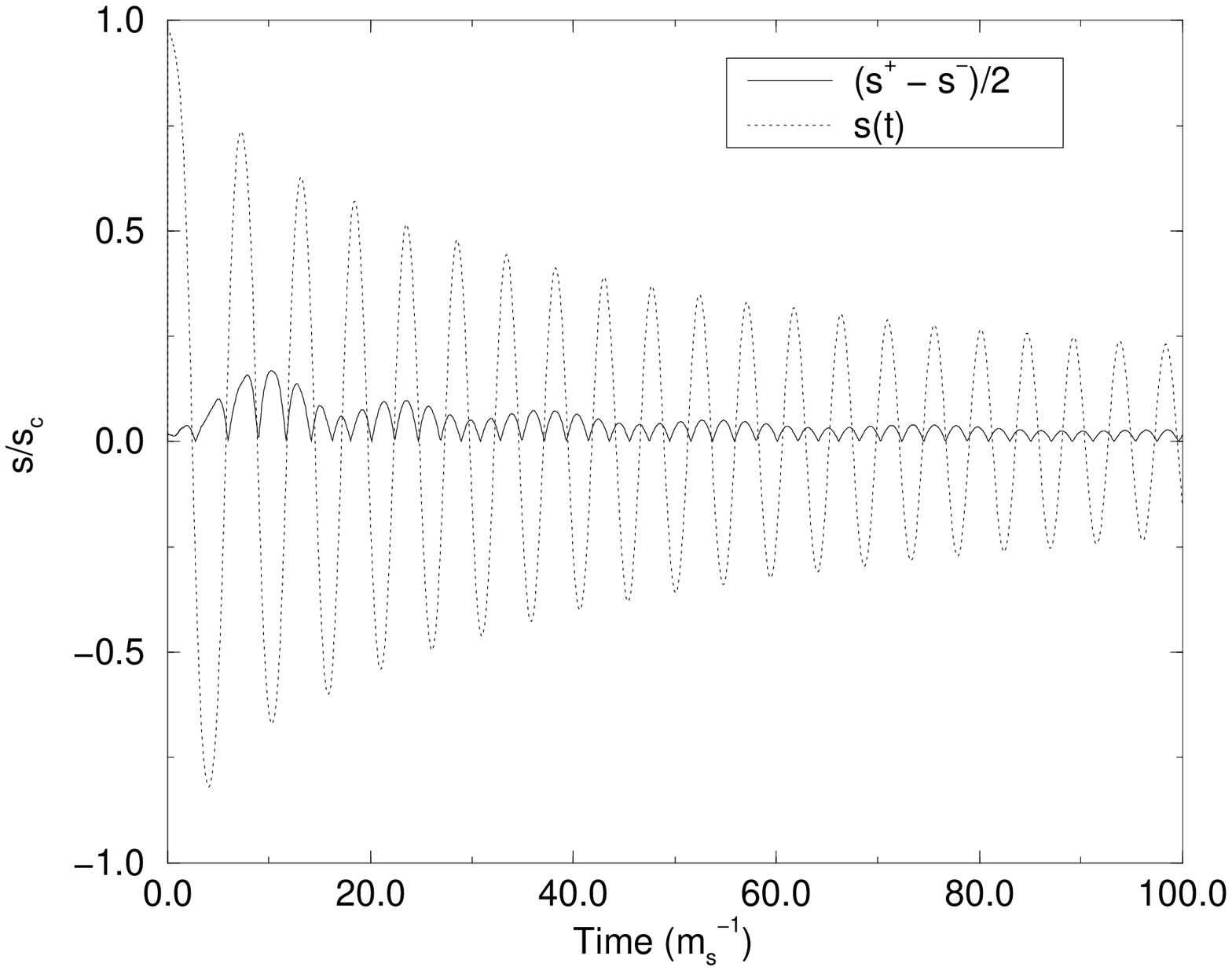}%
\setlength{\subFigureAboveCaptionSpace}{7mm} 
\end{arrangedFigure} 
\setlength{\subFigureAboveCaptionSpace}{0mm} 
\newpage

\section{Consequences of Inflatonic Non-Topological Soliton Formation} 

     In this section we discuss some possible cosmological 
consequences of non-topological soliton (NTS) 
formation. After fragmentation the inflaton sector fields are 
expected to form NTS which we have labelled  
inflaton condensate lumps \cite{icf}, with essentially 
all of the energy density of the Universe 
concentrated in the lumps \cite{icf,cpr}. Physically these are 
interpreted as a Bose condensate of 
scalars held together by an attractive self-interaction 
which balances against the gradient  
terms (uncertainty principle pressure) in the scalar field equations. The 
most obvious physical consequences of inflaton condensate lump 
formation will be related to the 
inhomogeneous distribution of the energy density of the Universe in the post-inflation 
NTS era and the inhomogeneous nature of reheating via NTS decay. 

      The decay rate of the condensate lumps is related to (i) its stability as a 
solution of the classical equations of motion and (ii) the decay and annihilation 
rate of the scalars inside the NTS. 
For the case of a real scalar field it is likely that the condensate lumps are not stable, 
but instead slowly radiate scalar field waves until (depending on the potential) the amplitude 
of the scalar field inside the lumps becomes too small for the attractive interaction term 
to hold the lumps together. 
This has been studied extensively in \cite{oscillons}, where the 
condensate lumps have been given the name 'oscillons'. The lifetime of the oscillons 
in a potential of the form $V(\phi) = m^{2}\phi^{2}/2 - \lambda \phi^{4}/4$ 
(which approximates D-term inflation \cite{icf}) has been estimated to be
around $10^{3-4}m^{-1}$ \cite{oscillons,iballs}.   
In this case we would expect the scalars forming the lump to eventually disperse.
However, during the period when the condensate lumps exist, the fact that the 
number density of scalars inside the lumps is constant (unlike the case of a 
conventional homogeneous inflaton condensate \cite{param2}) means that perturbative
inflaton annihilation and/or parametric resonance will be effectively enhanced 
and could dominate over perturbative inflaton decays 
as the main mode of reheating \cite{icf}. This all depends on 
whether the classical decay of the condensate lumps is slow enough relative to
 annihilation/parametric resonance for the condensate lumps 
to have non-trivial consequences for reheating.  

     However, an important point is that in SUSY hybrid inflation models
the inflaton is in fact a complex field. In D-term inflation there
is a global $U(1)_{S}$ symmetry 
such that $S \rightarrow e^{i \alpha}S$ and $\Phi_{+} \rightarrow e^{-i\alpha}\Phi_{+}$ 
in the superpotential. As a result, there will be a Q-ball solution of the inflaton sector 
equations of motion \cite{mattq} carrying a non-zero $U(1)_{S}$ charge. 
In this case it is likely that 
the initially neutral condensate lump will fragment into a 
$\pm$Q-ball pair \cite{enqkas}. This occurs 
because although initially the 
inflaton field and condensate lump is real, 
small fluctuations of the phase of the inflaton field inside the lump 
can grow through mode-mode couplings, resulting in a seperation of charge 
in the quasi-stable condensate lump into a Q-ball, anti-Q-ball pair. 
This has been demonstrated in lattice simulations 
of a running mass chaotic inflation model \cite{enqkas}. Although the potential 
of this single-field inflation model is different from that of D-term inflation, 
we would expect the tendency of the system to reach its most stable configuration 
($\pm$Q-ball pairs) in the presence of spatial perturbations to be repeated. 

     Since the Q-balls are classically stable,  
reheating will be determined by the decay/annihilations of the globally charged 
inflaton sector fields forming the Q-balls i.e. reheating will occur via 
Q-ball decay, with the energy density of the Universe 
concentrated inside the classically stable 
Q-balls until that time. Therefore reheating and thermalisation will be highly 
inhomogeneous in this case, unlike the case of a conventional homogeneous inflaton
 condensate.

     The formation of condensate lumps and fragmentation to Q-balls also 
results in a highly inhomogeneous energy density prior to reheating. 
One consequence of this is that 
the dynamics of scalar fields in SUSY and SUGRA cosmology will be altered.
In SUGRA models, Planck-scale suppressed corrections result in a 
correction to the mass squared term of scalar fields the form 
$|F|^{2}/M_{Pl}^{2} \approx \rho/M_{Pl}^{2}$, where $F$ represents the 
F-term of any field in the cosmological background and $\rho$ is its 
the energy density. In the conventional case of a homogeneous 
coherently oscillating inflaton field following inflation, $\rho$ is also homogeneous 
and so the correction to the mass squared term is of the form 
$\overline{\rho}/M_{Pl}^{2} \sim c H^{2}$, where $\overline{\rho}$ the average energy density and $|c|$ is 
of the order of 1 \cite{eta}. However, when the energy density is packed into Q-balls, 
the correction to the mass squared term of the scalars outside the Q-balls will 
be effectively zero, whilst inside it will be much larger than $H^{2}$. 
As a result, the dynamics of flat direction scalars in the MSSM \cite{enqmaz} 
(and other so-called moduli fields) 
will be completely different from the conventional case. In the bulk outside the Q-balls 
(corresponding to most of the volume of the Universe) the scalar fields will not evolve until 
$H^{2} \approx m_{susy}^{2}$ (which occurs prior to reheating if the reheating temperature satisfies 
the thermal gravitino constraint $T_{R} \lae 10^{8} \GeV$ \cite{grav}), 
where $m_{susy}^{2} \approx (100 \GeV)^{2}$ 
is the conventional gravity-mediated SUSY-breaking mass squared term. Inside or close to the Q-balls, if we consider 
the mass squared correction due to $\rho$ to be positive, the flat direction scalar will be 
driven towards zero. The full dynamics of the MSSM flat directions will require a complete 
solution of the scalar field equations in the presence of Q-balls, but we expect the dynamics in the bulk to be 
essentially that of $c = 0$ in the conventional
 discussion of MSSM flat directions and moduli. This will
have significant consequences for 
the cosmology of the MSSM, in particular for Affleck-Dine baryogenesis \cite{ad,enqmaz}.    
We expect other consequences of the NTS dominated post-inflation era 
to become apparent in the future.

\section{Conclusions} 

    We have analysed the growth of spatial perturbations of the inflaton sector  fields of 
D-term hybrid inflation including 1-loop radiative corrections, 
using a semi-analytic model with simplified initial conditions. 
We have argued that this model will give a good estimate of the growth of spatial
perturbation modes in the fully realistic case. 

     We find that the growth of the perturbations 
to fully non-linearity (indicating the onset of inflaton 
condensate lump formation) occurs once 
$\lambda/g \gae 0.09$, and that the maximal growth factor of the 
perturbations is a function only of $\lambda/g$. Therefore non-topological 
soliton formation is a natural possibility in D-term inflation models. 
The corresponding perturbation mode, which sets the length 
scale for the condensate lumps in the case where $\lambda/g \approx 0.09$, 
is given by $k \approx 
0.008s_{c}$ for $g =1$, with $k \propto g$ for fixed $\lambda/g$.  
The growth of spatial perturbations to 
full non-linearity is rapid, typically after about one coherent 
oscillation of the inflaton field, which is consistent with a tachyonic preheating 
picture \cite{tp} of the growth of spatial perturbations. However, the 
value of $k_{max}$, the largest wavenumber for which perturbation growth can 
occur, and the dependence of the fragmentation condition only on the ratio 
$\lambda/g$ are both consistent 
with the analytical predictions of the inflaton condensate 
fragmentation analysis \cite{icf}.  

    Although our analysis strictly applies to the case of D-term inflation, it
 strongly suggests that inflaton condensate fragmentation 
will be an unavoidable feature of F-term hybrid inflation, 
which has the same classical scalar potential as D-term inflation in the limit 
$\lambda = \sqrt{2} g$. However, a full analysis including the 
1-loop potential of F-term inflation remains to be done. 

     The formation of non-topological solitons (inflaton condensate lumps) is 
undoubtedly a feature of the cosmology of SUSY hybrid inflation models for
a natural range of the inflaton sector couplings, resulting 
in a highly inhomogeneous post-inflation era with all 
the energy of the Universe packed inside the non-topological 
solitons.  We have discussed some
possible consequences of inflaton condensate lump formation 
for SUSY cosmology, including the 
 formation of  inflatonic Q-balls from decay of the 
condensate lumps,
 inhomogeneous reheating and the modification of
 flat direction scalar field dynamics. We hope to develop 
a more detailed understanding of 
this post-inflation era in the future.  

\section*{Acknowledgements}

      The research of M.B. was supported by the EPSRC.

\end{document}